\documentclass[nofootinbib,twocolumn]{revtex4-1}

\usepackage[utf8]{inputenc}
\usepackage[T1]{fontenc}

\usepackage[dvipsnames]{xcolor}
\usepackage{hyperref}
\hypersetup{
  breaklinks=true,
  colorlinks=true,
  allcolors=BlueViolet,
}

\usepackage{physics,braket,bm} 
\renewcommand{\t}{\text} 
\newcommand{\f}[2]{\dfrac{#1}{#2}} 
\newcommand{\p}[1]{\left(#1\right)} 
\renewcommand{\sp}[1]{\left[#1\right]} 
\newcommand{\bk}{\Braket} 
\newcommand{\uv}[1]{\bm{\hat{#1}}} 
\renewcommand{\i}{\mathrm{i}\mkern1mu} 

\renewcommand{\set}[1]{\{#1\}} 
\renewcommand{\Set}[1]{\left\{#1\right\}} 

\newcommand{\bbk}[1]{\langle\!\langle #1 \rangle\!\rangle}
\newcommand{\Bbk}[1]
{\left\langle\!\!\left\langle #1 \right\rangle\!\!\right\rangle}

\newcommand{\x}{\text{x}}
\newcommand{\y}{\text{y}}
\newcommand{\z}{\text{z}}

\newcommand{\C}{\mathcal{C}}
\newcommand{\D}{\mathcal{D}}
\newcommand{\E}{\mathcal{E}}
\newcommand{\I}{\mathcal{I}}
\newcommand{\N}{\mathcal{N}}
\renewcommand{\O}{\mathcal{O}}
\newcommand{\Q}{\mathcal{Q}}
\renewcommand{\S}{\mathcal{S}}

\def\obra#1{\mathinner{({#1}|}}
\def\oket#1{\mathinner{|{#1})}}
\def\obk#1{\mathinner{({#1})}}

\DeclareMathOperator{\cov}{cov}
\DeclareMathOperator{\diag}{diag}

\usepackage{graphicx} 

\usepackage[inline]{enumitem} 
\setlist[enumerate,1]{label={(\roman*)}} 

\usepackage{color} 

\begin{document}

\title{Spin qudit tomography and state reconstruction error}
\author{Michael A.~Perlin}
\email{mika.perlin@gmail.com}
\author{Diego Barberena}
\author{Ana Maria Rey}
\affiliation{JILA, National Institute of Standards and Technology and University of Colorado, 440 UCB, Boulder, Colorado 80309, USA}
\affiliation{Center for Theory of Quantum Matter, University of Colorado, Boulder, CO, 80309, USA}


\begin{abstract}
We consider the task of performing quantum state tomography on a $d$-level spin qudit, using only measurements of spin projection onto different quantization axes.
After introducing a basis of operators closely related to the spherical harmonics, which obey the rotational symmetries of spin qudits, we map our quantum tomography task onto the classical problem of signal recovery on the sphere.
We then provide algorithms with $O\p{rd^3}$ serial runtime, parallelizable down to $O\p{rd^2}$, for (i) computing a priori upper bounds on the expected error with which spin projection measurements along $r$ given axes can reconstruct an unknown qudit state, and (ii) estimating a posteriori the statistical error in a reconstructed state.
Our algorithms motivate a simple randomized tomography protocol, for which we find that using more measurement axes can yield substantial benefits that plateau after $r\approx3d$.
\end{abstract}

\maketitle

\section{Introduction}

Quantum state tomography, the task of reconstructing a quantum state by collecting and processing measurement data, is an essential primitive for quantum sensing, quantum simulation, and quantum information processing.
The central importance of quantum state tomography has led to the development of techniques based on least-squares inversion \cite{opatrny1997leastsquares}, linear regression \cite{qi2013quantum}, maximum likelihood estimation \cite{teo2011quantumstate, smolin2012efficient}, Bayesian inference \cite{huszar2012adaptive, ferrie2014quantum, granade2016practical}, compressed sensing \cite{gross2010quantum, kalev2015quantum}, and neural networks \cite{torlai2018neuralnetwork}, among others.
These techniques are typically developed in a general, information-theoretic setting, and make minimal assumptions about the physical medium of a quantum state.
As a consequence, even well-established techniques can be ill-suited for physical platforms with unique or limited capabilities.

Due to advancements in experimental capabilities to address nuclear spin states (i.e.~hyperfine levels) in ultracold atomic systems \cite{daley2011quantum, lu2011strongly, mischuck2012control, aikawa2012boseeinstein, smith2013quantum, cazalilla2014ultracold}, as well as developments in the control of ultracold molecular systems \cite{bohn2017cold, takekoshi2014ultracold, zeppenfeld2012sisyphus, kozyryev2017sisyphus, puri2017synthesis, wu2017cryofuge, marco2019degenerate, liu2019molecular, anderegg2019optical, chou2020frequencycomb, lin2020quantum}, a particular setting of growing interest is the spin qudit, or a multilevel quantum angular momentum degree of freedom.
Spin qudits can provide advantages over their qubit counterparts for quantum sensing \cite{hemmer2018squeezing, evrard2019enhanced}, enable quantum simulations of SU($d$) magnetism \cite{cazalilla2014ultracold, banerjee2013atomic, zhang2014spectroscopic, scazza2014observation, goban2018emergence, perlin2021engineering}, and offer unique capabilities for quantum computation and error correction \cite{albert2020robust, gross2021designing, barnes2021assembly}.
In all cases, quantum state tomography is necessary to take full advantage of a spin qudit\footnote{Note that the measurement of collective observables for quantum sensing or simulation can be recast as a single-spin tomography task.}.

The problem of qudit tomography is not new, with an extensive literature on a variety of techniques \cite{newton1968measurability, hofmann2004quantumstate, filippov2010inverse, schmied2011tomographic, evrard2019enhanced, flammia2005minimal, thew2002qudit, salazar2012quantum, sosa-martinez2017quantum, ha2018minimal, stefano2019set, palici2020oam}.
However, most existing protocols either rely on infinite-dimensional representations of a quantum spin \cite{manko1997spin, schmied2011tomographic, evrard2019enhanced}, or require the capability to perform essentially arbitrary operations on a qudit \cite{thew2002qudit, flammia2005minimal, salazar2012quantum, sosa-martinez2017quantum, ha2018minimal, stefano2019set, palici2020oam}, generally resulting in tomographic protocols that can be highly inefficient or unachievable in practice.
The protocols based on infinite-dimensional representations of a quantum spin have the advantage of reconstructing its state from measurements of spin projection onto different spatial axes, which are generally accessible with any spin qudit.
Nonetheless, these protocols obfuscate the minimal requirements for performing full state tomography, provide no straightforward error bounds or guarantees of accuracy, and (with the notable exception of Ref.~\cite{schmied2011tomographic}) generally extract only a small fraction of the information contained in measurement data.

In this work, we consider the task of performing spin qubit tomography using only measurements of spin projection onto different spatial axes.
This sort of task was first considered in Ref.~\cite{newton1968measurability}, as well as a few later works \cite{hofmann2004quantumstate, filippov2010inverse, schmied2011tomographic}.
Specifically, Ref.~\cite{newton1968measurability} provided an explicit protocol for reconstructing a $d$-level spin qudit state from measurements of spin projection along $2d-1$ axes, the minimum number necessary for full tomography of an arbitrary (possibly mixed) qudit state.
However, the protocol in Ref.~\cite{newton1968measurability} involves a choice of a single (arbitrary) angle $\theta$, and provides no means for comparing different choices of $\theta$, which may result in wildly different statistical errors (i.e.~precision) in a reconstructed state.
Other works provide insightful discussions into the problem of spin qudit tomography, but either
\begin{enumerate*}
\item require making assumptions about the qudit state in question \cite{schmied2011tomographic} (making the tomographic protocol only valid for a restricted set of possible states),
\item do not address the question of statistical error \cite{hofmann2004quantumstate}, or
\item provide a measure of statistical error that is needlessly conservative and computationally demanding \cite{filippov2010inverse}.
\end{enumerate*}
We address these shortcomings in this work, and identify remaining avenues for refining spin qudit tomography protocols.

In Section \ref{sec:polarization_ops}, we introduce a set of qudit operators that are closely related to the spherical harmonics, and which play a central role in our work.
We then map the quantum problem of spin qudit tomography onto the classical problem of signal recovery on the sphere in Section \ref{sec:signal_recovery}, thereby providing an intuitive perspective on spin qudit tomography.
In Section \ref{sec:error} we provide a priori upper bounds and a posteriori estimates of the statistical error in a qudit state reconstructed from measurements of spin projection along a given set of $r$ measurement axes.
The capability to determine upper bounds on reconstruction error a priori motivates a simple randomized tomography protocol that we outline in Section \ref{sec:protocol}, and for which we numerically find that using more measurement axes yields substantial benefits that plateau after $r\approx3d$.
To facilitate the use of our protocols, we make all of our codes publicly available at Ref.~\cite{tomo_codes}, which also contains the best measurement axes we found for $d\le30$ and $r=3d$.

\section{Polarization operators}
\label{sec:polarization_ops}

We begin by introducing a set of qudit operators that are closely related to the spherical harmonics (in a sense that will be clarified below), and which play a central role in our work.
Consider a $d$-state spin qudit with total spin $s\equiv\frac{d-1}{2}$.
The defining property of a spin qudit, distinguishing it from other qudits, is that it describes an angular momentum degree of freedom, which has specific implications for how a spin qudit should transform under the group SO(3) of rotations in 3D space.
Due to the central importance of these transformation rules for a spin qudit, we seek a basis of operators that transform nicely under 3D rotations\footnote{Technically speaking, we seek a basis of operators that transform as an irreducible representation of SO(3).}.
One such basis is that of the {\it polarization operators} \cite{kryszewski2006positivity, bertlmann2008bloch}, defined by
\begin{align}
  T_{\ell m} \equiv \sqrt{\f{2\ell+1}{2s+1}} \sum_{\mu,\nu=-s}^s
  \bk{s\mu;\ell m|s\nu} \op{\nu}{\mu},
  \label{eq:trans_op}
\end{align}
where $\ket\mu$ is an eigenstate of the axial spin projection operator $S_\z\ket\mu=\mu\ket\mu$; and $\bk{s\mu;\ell m|s\nu}$ is a Clebsh-Gordan coefficient that enforces $\ell\in\set{0,1,\cdots,d-1}$ and $m\in\set{-\ell,-\ell+1,\cdots,\ell}$, such that there are $d^2$ polarization operators in total.
For brevity, we will generally treat the value of $d$ as constant but arbitrary throughout this work, and we will suppress any explicit dependence of quantities or operators such as $T_{\ell m}$ on $d$.
The polarization operators are orthonormal with respect to the trace inner product, and transform nicely under conjugation:
\begin{align}
  \obk{T_{\ell m }|T_{\ell'm'}}
  = \delta_{\ell\ell'} \delta_{mm'},
  &&
  T_{\ell m}^\dag = \p{-1}^m T_{\ell,-m},
\end{align}
where for any $d\times d$ matrix $X=\sum_{\mu,\nu} X_{\mu\nu} \op{\mu}{\nu}$ we define the $d^2$-component vector $\oket{X} \equiv \sum_{\mu,\nu} X_{\mu\nu} \ket{\mu\nu}$; $\obra{X}$ is the conjugate transpose of $\oket{X}$, such that $\obk{X|Y}=\tr\p{X^\dag Y}$; and $\delta_{kk'}\equiv 1$ if $k=k'$ and $0$ otherwise.
These properties of the polarization operators allow us to expand any density operator $\rho$ in the polarization operator basis as
\begin{align}
  \rho = \sum_{\ell=0}^{d-1} \sum_{m=-\ell}^\ell
  \rho_{\ell m} T_{\ell m},
  &&
  \rho_{\ell m} \equiv \bk{T_{\ell m}^\dag}_\rho,
  \label{eq:trans_state}
\end{align}
where $\bk{X}_\rho\equiv\tr\p{\rho X}=\obk{\rho|X}$, and $\rho^\dag=\rho$ implies that $\rho_{\ell m}^*=\p{-1}^m\rho_{\ell,-m}$.
The polarization operators can be interpreted in terms of an absorption process, whereby $T_{\ell m}\ket\psi$ is (up to normalization) the state obtained after a spin-$s$ state $\ket\psi$ absorbs a particle with total spin $\ell$ and spin projection $m$ onto a fixed quantization axis.
Similarly to the complex spherical harmonics $Y_{\ell m}$, we will refer to $\ell$ as the {\it degree} and $m$ as the {\it order} of $T_{\ell m}$.

The polarization operators are spherical tensor operators, whose degree is preserved under 3D rotations generated by the spin operators $S_\x,S_\y,S_\z$.
Moreover, the degree-$\ell$ polarization operators $T_{\ell m}$ transform similarly to spin-$\ell$ particles and spherical harmonics $Y_{\ell m}$ under 3D rotations (see Appendix \ref{sec:rotations}).
Specifically, for any triplet of angles $\bm\omega=\p{\alpha,\beta,\gamma}$, we can therefore define the rotation operator
\begin{align}
  R\p{\bm\omega} \equiv e^{-\i\alpha S_\z} e^{-\i\beta S_\y} e^{-\i\gamma S_\z},
\end{align}
and expand rotated polarization operators as
\begin{align}
  T_{\bm\omega\ell m} \equiv
  R\p{\bm\omega} T_{\ell m} R\p{\bm\omega}^\dag
  = \sum_{n=-\ell}^\ell D_{mn}^\ell\p{\bar{\bm\omega}}^* T_{\ell n},
  \label{eq:trans_rot}
\end{align}
where $\bar{\bm\omega}=\p{\gamma,\beta,\alpha}$ is the reversal of $\bm\omega$, and
\begin{align}
  D_{mn}^\ell\p{\bar{\bm\omega}}
  \equiv \bk{\ell m|R\p{\bar{\bm\omega}}|\ell n}
  \label{eq:wigner_D}
\end{align}
are (Wigner) rotation matrix elements.
For reasons that will become clear shortly, throughout this work we will primarily consider rotations of the sphere that take the north pole to a point $\bm v=\p{\alpha,\beta}$ at azimuthal angle $\alpha$ and polar angle $\beta$.
For ease of notation, we therefore define $R\p{\bm v} \equiv R\p{\alpha,\beta,0}$, $T_{\bm v\ell m} \equiv T_{\p{\alpha,\beta,0},\ell m}$, and $D^\ell_{mn}\p{\bm v} \equiv D^\ell_{mn}\p{0,\beta,\alpha}$.

The polarization operators $T_{\ell m}$ share a connection to the spherical harmonics $Y_{\ell m}$ that goes beyond the rules for their transformation under 3D rotations.
In fact, the phase-space representation of $T_{\ell m}$ is proportional to $Y_{\ell m}$.
The phase-space representation of a spin qudit operator $X$ assigns, to each point $\bm v$ on the sphere, the complex number
\begin{align}
  X^{\t{PS}}\p{\bm v} \equiv \bk{s_{\bm v}|X|s_{\bm v}},
\end{align}
where $\ket{s_{\bm v}}\equiv R\p{\bm v}\ket{s}$ is the state of a spin qudit polarized along $\bm v$.
This representation is faithful in the sense that $X$ is uniquely determined by the phase-space values $X^{\t{PS}}\p{\bm v}$ at all points $\bm v$ on the sphere.
The transformation rules for polarization operators in Eq.~\eqref{eq:trans_rot}, together with the fact that $\bk{s|T_{\ell m}|s}=0$ unless $m=0$, suffice to show that
\begin{align}
  T_{\ell m}^{\t{PS}}\p{\bm v} = c_\ell Y_{\ell m}\p{\bm v},
\end{align}
where the scalar $c_\ell$ simply enforces $\obk{T_{\ell m}|T_{\ell m}}=1$ (see Appendix \ref{sec:rotations}).
The polarization operators $T_{\ell m}$ are thus a quantum analogue of the spherical harmonics $Y_{\ell m}$, and play an important role in phase-space formalisms for spin qudits \cite{li2013weylwignermoyal}.

As a special case, the phase-space representation $\rho^{\t{PS}}$ of a spin qudit state $\rho$ is commonly known as its Husimi distribution.
Performing tomography on an unknown qudit state $\rho$ is therefore equivalent to reconstructing the unknown distribution $\rho^{\t{PS}}$ on the sphere.
In principle, the representation $\rho^{\t{PS}}$ of a finite-dimensional qudit state $\rho$ can be reconstructed from the values $\rho^{\t{PS}}\p{\bm v}=\bk{s_{\bm v}|\rho|s_{\bm v}}$ at a finite number of points $\bm v$.
In practice, the value $\bk{s_{\bm v}|\rho|s_{\bm v}}$ is determined by measuring spin projection along $\bm v$, which also provides measurement data on all spin projections $\bk{\mu_{\bm v}|\rho|\mu_{\bm v}}$ with $\mu\in\set{s,s-1,\cdots,-s}$ and $\ket{\mu_{\bm v}}\equiv R\p{\bm v}\ket{\mu}$; one would like to make use of this additional data as well.
We clarify the connection between the quantum problem of reconstructing $\rho$ from spin projection measurements and the classical problem of reconstructing $\rho^{\t{PS}}$ from its values $\rho^{\t{PS}}\p{\bm v}$ in the following section.

\section{Spin tomography as signal recovery on the sphere}
\label{sec:signal_recovery}

Our goal is to reconstruct an arbitrary state $\rho$ of a spin qudit from measurements of spin projection onto different quantization axes.
We are thus nominally restricted to measuring projectors $\Pi_{\bm v\mu} \equiv \op{\mu_{\bm v}}$, where $\ket{\mu_{\bm v}}\equiv R\p{\bm v}\ket{\mu}$ is a state with spin projection $\mu$ onto the measurement axis $\bm v$.
For any fixed axis $\bm v$, the sets $\set{\Pi_{\bm v\mu}}$ and $\set{T_{\bm v\ell,0}}$ (i.e.~all $T_{\bm v\ell m}$ with $m=0$) are both complete bases for the space of operators that are diagonal in the basis $\set{\ket{\mu_{\bm v}}}$.
Measuring the projectors $\set{\Pi_{\bm v\mu}}$ is therefore equivalent to measuring the polarization operators $\set{T_{\bm v\ell,0}}$, and provides data on the expectation values $\bk{T_{\bm v\ell,0}}_\rho$.

In order to reconstruct an arbitrary density operator $\rho$ from the expectation values $\bk{T_{\bm v\ell,0}}_\rho$, we essentially need to find a set of coefficients $C_{\ell mk}\p{\bm v}$ that would allow us to recover any matrix element $\rho_{\ell m}$ of $\rho$ through
\begin{align}
  \rho_{\ell m}^* = \bk{T_{\ell m}}_\rho
  = \sum_{\bm v,k} C_{\ell mk}\p{\bm v} \bk{T_{\bm vk,0}}_\rho.
  \label{eq:state_recon}
\end{align}
Expanding the rotated polarization operators $T_{\bm vk,0}$ into a sum of un-rotated polarization operators $T_{\ell n}$ according to Eq.~\eqref{eq:trans_rot}, we find that the recovery condition in Eq.~\eqref{eq:state_recon} is satisfied when
\begin{align}
  T_{\ell m}
  = \sum_{\bm v,k,n} C_{\ell mk}\p{\bm v} D^k_{0,n}\p{\bm v}^* T_{kn}.
\end{align}
Orthogonality of the polarization operators then implies the decomposition $C_{\ell mk}\p{\bm v}=\delta_{\ell k}C_{\ell m}\p{\bm v}$, and in turn
\begin{align}
  \sum_{\bm v} C_{\ell m}\p{\bm v} D^\ell_{0,n}\p{\bm v}^*
  = \delta_{mn}
  \label{eq:tomo_recovery}
\end{align}
for all $\ell$.

\begin{figure}
  \centering
  \includegraphics[width=0.4\columnwidth]{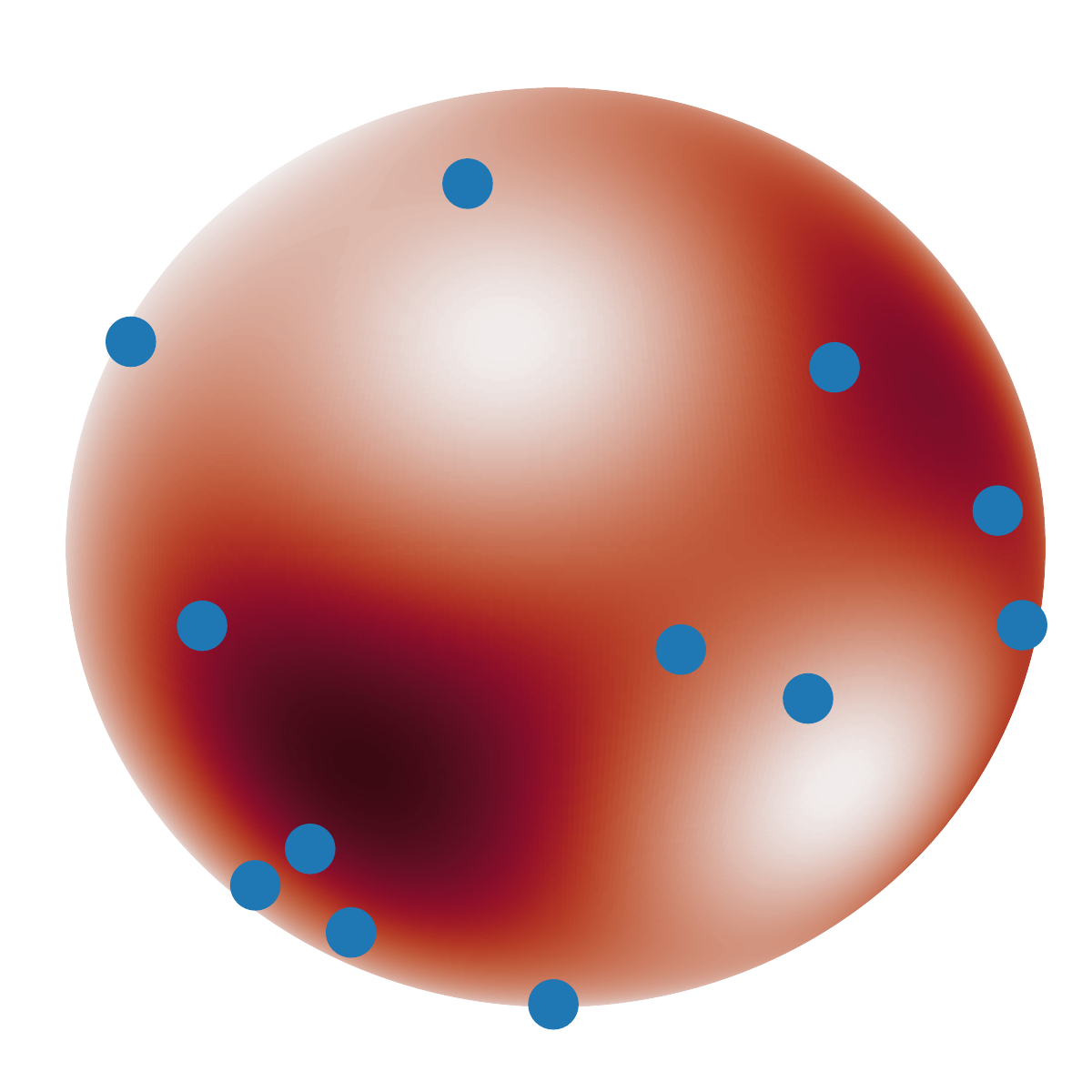}
  \caption{Signal recovery on the sphere is the problem of reconstructing an unknown function $f$ (red distribution) from its values $f\p{\bm v}$ at specific points $\bm v\in V$ (blue dots) on the sphere.
    For almost all choices of $V$, reconstruction of $f$ is possible if there are at least as many points in $V$ as there are degrees of freedom in $f$.}
  \label{fig:sphere_points}
\end{figure}

In fact, the problem of finding suitable axes $V$ and coefficients $C_{\ell m}\p{\bm v}$ to satisfy Eq.~\eqref{eq:tomo_recovery} can be mapped onto the well-studied problem of signal recovery on the sphere (see Figure \ref{fig:sphere_points}) \cite{mcewen2011novel, rauhut2011sparse, alem2012sparse, khalid2014optimaldimensionality}.
The signal recovery problem can be stated as follows: given a square-integrable function $f$ on the sphere, with the spherical harmonic expansion
\begin{align}
  f\p{\bm v} = \sum_{\ell,m} f_{\ell m} Y_{\ell m}\p{\bm v},
\end{align}
where $f_{\ell m}$ are complex coefficients, find a set of points $V=\set{\bm v}$ and associated coefficients $\tilde C_{\ell m}\p{\bm v}$ with which we can reconstruct $f$, or equivalently its coefficients $f_{\ell m}$, from knowledge of the function's value $f\p{\bm v}$ at all points $\bm v\in V$; that is
\begin{align}
  f_{\ell m} = \sum_{\bm v} \tilde C_{\ell m}\p{\bm v} f\p{\bm v}
  = \sum_{\bm v,k,n} \tilde C_{\ell m}\p{\bm v} Y_{kn}\p{\bm v} f_{kn}.
\end{align}
Reconstruction of functions with arbitrary coefficients $f_{\ell m}$ implies that
\begin{align}
  \sum_{\bm v} \tilde C_{\ell m}\p{\bm v} Y_{kn}\p{\bm v}
  = \delta_{\ell k} \delta_{mn},
  \label{eq:full_recovery}
\end{align}
which is a stronger version of the condition that we found for the spin qudit tomography problem in Eq.~\eqref{eq:tomo_recovery}.
We will refer to Eq.~\eqref{eq:full_recovery} as the {\it full recovery problem}, and Eq.~\eqref{eq:tomo_recovery} as the the {\it reduced recovery problem}.
Due to the fact that $D^\ell_{0,m}\p{\bm v} = \sqrt{\frac{4\pi}{2\ell+1}}\, Y_{\ell m}\p{\bm v}$, any solution to the full recovery problem automatically solves the reduced recovery problem by setting $C_{\ell m}\p{\bm v} = \sqrt{\frac{2\ell+1}{4\pi}}\, \tilde C_{\ell m}\p{\bm v}^*$.
In principle, this mapping allows us to import a host of existing signal recovery algorithms \cite{mcewen2011novel, rauhut2011sparse, alem2012sparse, khalid2014optimaldimensionality} for the task of spin qudit tomography.
In practice, spin qudits typically have only a modest dimension $d$, which allows for simpler and optimized tomography protocols that are practical despite worse scaling with $d$ (see Section \ref{sec:protocol}).
A natural avenue to develop better spin qudit tomography protocols would therefore be to build on the existing classical signal recovery algorithms, tailoring them to solve the reduced recovery problem in Eq.~\eqref{eq:tomo_recovery} rather than the full recovery problem in Eq.~\eqref{eq:full_recovery}.
We leave these developments to future work.

If the function $f$ is {\it band-limited} at degree $L$, which is to say that $f_{\ell m}=0$ for all $\ell\ge L$, then the full recovery problem in Eq.~\eqref{eq:full_recovery} is provably solvable with a suitable choice of $\abs{V}=L^2$ points on the sphere \cite{freeden2008spherical, freeden2018spherical}.
The existence of these solutions to the full recovery problem in turn implies the existence of $d^2$ measurement axes that suffice to reconstruct arbitrary states of $d$-level spin qudit, whose possible states (or rather, phase-space representations) are band-limited at degree $d$.
Moreover, for any fixed degree $\ell$, finding solutions to the reduced recovery problem in Eq.~\eqref{eq:tomo_recovery} is equivalent to the recovery of a degree-$\ell$ function $f_\ell = \sum_m f_{\ell m} Y_{\ell m}$, which is provably possible with $\abs{V}=2\ell+1$ samples \cite{freeden2008spherical}.
In the case of spin qudit tomography, the degree $\ell$ takes a maximal value of $\ell_{\t{max}}\equiv d-1$, so state recovery requires as many measurement axes as there are polarization operators with degree $\ell_{\t{max}}$, namely $2\ell_{\t{max}}+1=2d-1$.

\section{State reconstruction error}
\label{sec:error}

For the practically minded, proving the existence of solutions to a problem is less interesting than the exposition of a particular solution.
On a high level, a spin qudit tomography protocol consists of
\begin{enumerate*}
\item selecting a set of measurement axes,
\item \label{tomo:measure} collecting measurement data on spin projection onto these axes, and then
\item processing the collected data to reconstruct the state of the spin qudit.
\end{enumerate*}
Whereas step \ref{tomo:measure} can involve a host of platform-dependent technical challenges, in the following sections we discuss the steps to take before and after collecting measurement data.

To this end, we begin by asking a question: what is a ``good'' choice of measurement axes?
Intuitively, a good choice of axes should minimize the error with which one can reconstruct an unknown quantum state from associated measurement data.
If we can quantify this intuition, then we can optimize over different choices of measurement axes to find a set that (approximately) minimizes the error in reconstructed states.

A set of measurement axes $V=\set{\bm v}$ nominally induces a set of projectors $\set{\Pi_{\bm v\mu}}$ that will be measured in an experiment.
By a simple change of basis, measuring these projectors is equivalent to measuring the polarization operators $\set{T_{\bm v\ell,0}}$.
Flattening each $d\times d$ matrix $T_{\bm v\ell,0}$ into the $d^2$-component column vector $\oket{T_{\bm v\ell,0}}$, we construct the {\it measurement matrix}
\begin{align}
  M_V \equiv \sum_{\bm v,\ell} \ket{\bm v\ell} \obra{T_{\bm v\ell,0}}.
  \label{eq:meas_mat}
\end{align}
Here $\bm v$ and $\ell$ label a row of $M_V$, or equivalently label a standard (``one-hot'') basis vector $\ket{\bm v\ell}$ of a $\p{\abs{V}\times d}$-dimensional vector space, and $\obra{T_{\bm v\ell,0}}$ is the conjugate transpose of $\oket{T_{\bm v\ell,0}}$.
A necessary and sufficient condition for $V$ to allow for full state tomography is that the measured polarization operators $T_{\bm v\ell,0}$, or equivalently the rows of $M_V$, span the entire ($d^2$-dimensional) space of operators on a $d$-level spin qudit.
In this case $M_V$ must be full rank, with $d^2$ nonzero singular values.
Indexing these singular values $M^V_k$ and the corresponding (normalized) left singular vectors $\bm x^V_k\equiv \sum_j x^V_{kj} \ket{j}$ by an integer $k\in\set{1,2,\cdots,d^2}$, we can construct the orthonormal qudit operators
\begin{align}
  Q^V_k \equiv \sum_j \p{q^V_{kj}}^* T_j,
  &&
  q^V_{kj} \equiv \f{x^V_{kj}}{M^V_k},
\end{align}
where for shorthand we use a combined index $j=\p{\bm v,\ell}$ to specify both a measurement axis $\bm v$ and a degree $\ell$, which identify the polarization operator $T_j\equiv T_{\bm v\ell,0}$.
These operators allow us to expand any state $\rho$ of a $d$-level spin qudit in the form
\begin{align}
  \rho = \sum_{k=1}^{d^2} \rho_k^V Q_k^V,
  &&
  \rho_k^V \equiv \bk{{Q^V_k}^\dag}_\rho.
\end{align}
Given empirical estimates $\tilde T_j$ of the expectation values $\bk{T_j}_\rho$, an empirical estimate $\tilde\rho_V$ of $\rho$ is then
\begin{align}
  \tilde\rho_V \equiv \sum_k \tilde\rho^V_k Q^V_k,
  \label{eq:reconstructed_state}
\end{align}
where, using the fact that $T_j=T_j^\dag$ (because they are diagonal polarization operators with degree $m=0$),
\begin{align}
  \tilde\rho^V_k \equiv \sum_j q^V_{kj} \tilde T_j
  \approx \sum_j q^V_{kj} \bk{T_j}_\rho
  = \bk{{Q^V_k}^\dag}_\rho
  = \rho^V_k.
\end{align}
The measurement matrix $M_V$ allows us to make concrete statements about the statistical error between the empirical estimate $\tilde\rho_V$ and the true state $\rho$.
Assume, for example, that the estimates $\tilde T_j$ are equal to $\bk{T_j}_\rho$ up to uncorrelated noise with variance no grater than $\epsilon^2$:
\begin{align}
  \tilde T_j = \bk{T_j}_\rho + \epsilon_j,
  &&
  \bbk{\epsilon_j \epsilon_{j'}} \le \epsilon^2 \delta_{jj'}.
  \label{eq:error_assumption}
\end{align}
Here $\set{\epsilon_j}$ are independent random variables, and we use the double brackets $\bbk{\cdot}$ to denote statistical averaging over experimental trials that estimate $\bk{T_j}_\rho$.
In this case, the mean squared error with which $\tilde\rho^V_k$ approximates $\rho^V_k$ is
\begin{align}
  \Bbk{\abs{\tilde\rho^V_k-\rho^V_k}^2}
  &= \Bbk{\p{\tilde\rho^V_k-\rho^V_k}^*\p{\tilde\rho^V_k-\rho^V_k}} \\
  &= \sum_{j,j'} \p{q^V_{kj}}^* q^V_{kj'} \,
  \bbk{\epsilon_j \epsilon_{j'}} \\
  &\le \sum_j \abs{q^V_{kj}}^2 \epsilon^2
  = \p{\f{\epsilon}{M^V_k}}^2.
\end{align}
Using the fact that the operators $Q^V_k$ are orthonormal, we can therefore bound the mean squared (Euclidean) distance between $\tilde\rho_V$ and $\rho$ as
\begin{align}
  \E_V\p{\rho}^2
  \equiv \Bbk{\norm{\tilde\rho_V-\rho}^2}
  \le \epsilon^2 \S_V^2,
  \label{eq:bound_eps}
\end{align}
where $\norm{X}^2 \equiv \obk{X|X} = \tr\p{X^\dag X}$ is the squared (Euclidean, Frobenius, or Hilbert-Schmidt) norm of $X$, and the {\it classical error scale} $\S_V$ is defined by
\begin{align}
  \S_V^2 \equiv \sum_k \p{M^V_k}^{-2} = \norm{M_V^{-1}}^2,
\end{align}
where $M_V^{-1}$ is the left inverse of $M_V$, satisfying $M_V^{-1} M_V = 1$.
We refer to the error scale $\S_V$ as ``classical'' because the bound in Eq.~\eqref{eq:bound_eps} applies in the presence of classical sources of measurement error.
Note that the classical error scale $\S_V$ diverges if the measurement matrix $M_V$ is singular, which indicates that measuring spin projections along all axes in $V$ does not provide sufficient information to reconstruct arbitrary quantum states.

\begin{figure}
  \centering
  \includegraphics{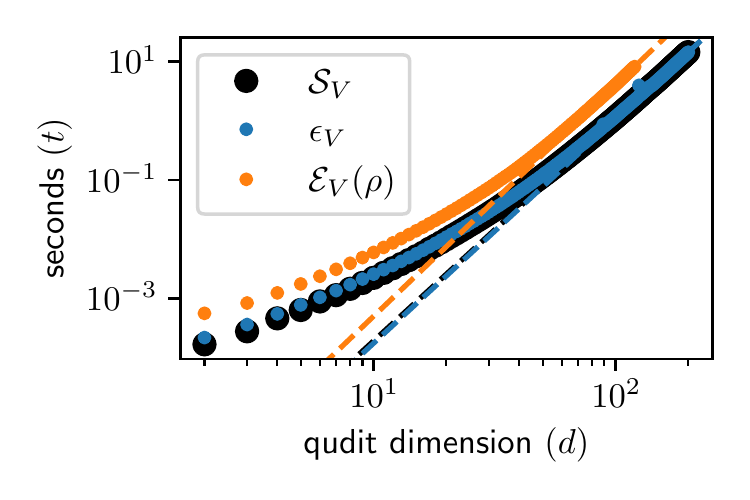}
  \caption{Serial runtime to compute $\S_V$, $\epsilon_V$, or $\E_V\p{\rho}$ with $\abs{V}=2d-1$ randomly chosen measurement axes and a randomly chosen qudit state $\rho$.
    Each point is an average over $10^3$ calculations or 5 minutes of runtime, whichever comes first.
    These results do not count fixed runtimes to pre-compute quantities that can be recycled for every new choice of $V$ and $\rho$.
    Dashed lines show fits to a runtime $t=c d^\alpha$ for the 20 largest values of $d$, finding $\alpha\approx 3.8\pm0.1$.}
  \label{fig:times}
\end{figure}

Computing the classical error scale $\S_V$ and estimates $\tilde\rho^V_k\approx\rho^V_k$ requires building the measurement matrix $M_V$ and computing its singular value decomposition.
The complexity of this task can be greatly reduced by the fact that the degree $\ell$ of a polarization operator $T_{\ell m}$ is preserved under rotations, which implies that the unitary
\begin{align}
  U \equiv \sum_{\ell=0}^{d-1} \sum_{m=-\ell}^\ell
  \oket{T_{\ell m}} \bra{\ell m},
  \label{eq:trans_basis}
\end{align}
with vectors $\oket{T_{\ell m}}$ in a column indexed by integers $\p{\ell,m}$, block-diagonalizes the measurement matrix into $d$ blocks indexed by the degree $\ell$:
\begin{align}
  M_V U = \sum_{\ell=0}^{d-1} \op{\ell} \otimes M_{V\ell},
\end{align}
where the $\abs{V}\times\p{2\ell+1}$-sized blocks are
\begin{align}
  M_{V\ell} \equiv \sum_{\bm v,m} \ket{\bm v}
  \obk{T_{\bm v\ell,0}|T_{\ell m}} \bra{m}
  = \sum_{\bm v,m} D^\ell_{0,m}\p{\bm v} \op{\bm v}{m}.
  \label{eq:block}
\end{align}
Here $D^\ell_{0,m}\p{\bm v}$ is a Wigner rotation matrix element, defined in Eq.~\eqref{eq:wigner_D}.
As the singular values of $M_V$ are invariant under unitary transformations, it follows that
\begin{align}
  \S_V^2 = \sum_\ell \S_{V\ell}^2,
  &&
  \S_{V\ell}^2 \equiv \norm{M_{V\ell}^{-1}}^2,
\end{align}
where $M_{V\ell}^{-1}$ is the left inverse of $M_{V\ell}$.
Constructing the block $M_{V\ell}$ and computing its singular value decomposition takes at most $O(\abs{V}d^2)$ time.
If we assume that $\abs{V}\sim d$, then computing the classical error scale $\S_V$ takes $O(d^4)$ serial or $O(d^3)$ parallel runtime (see Figure \ref{fig:times}).

The assumption that observables can be estimated up to uncorrelated noise with maximal variance $\epsilon^2$, summarized by Eq.~\eqref{eq:error_assumption}, is reasonable when measurement error is dominated by classical sources of experimental noise.
However, this assumption breaks down when measurement error is limited by fundamental quantum shot noise (i.e.~finite sampling error).
We relax the assumption of Eq.~\eqref{eq:error_assumption} in Appendix \ref{sec:bound}, where we instead assume that $\tilde\rho_V$ is built from $n$ independent measurements of spin projection along every axis $\bm v\in V$, with shot noise the dominant source of error.
In this case, the constraints that $\tr\p\rho=1$ and $\bk{\Pi_{\bm v\mu}}_\rho\ge0$ allow us to bound the mean squared distance between $\tilde\rho_V$ and $\rho$ as
\begin{align}
  \E_V\p{\rho}^2 < \f{\epsilon_V^2}{n},
  &&
  \epsilon_V^2 \equiv \sum_\ell \Gamma_\ell^2 \S_{V\ell}^2,
  \label{eq:bound}
\end{align}
where the {\it quantum error scale} $\epsilon_V$ is defined in terms of the spectral range of $T_{\ell,0}$:
\begin{align}
  \Gamma_\ell \equiv \f{\max_\mu t_{\ell\mu} - \min_\mu t_{\ell\mu}}{2},
  &&
  t_{\ell\mu} \equiv \bk{\mu|T_{\ell,0}|\mu}.
\end{align}
If $d$ is even or $\ell$ is odd, then $\Gamma_\ell=\max_\mu t_{\ell\mu}$.
For comparison with the ``classical'' error bound in Eq.~\eqref{eq:bound_eps}, we note that $\epsilon_V^2<\S_V^2/2$, so the previous bound still holds with the replacement $\epsilon^2\to1/2n$.
The factors $\Gamma_\ell^2$ are quick to compute and can be recycled for every new choice of axes $V$, so the complexity of computing $\epsilon_V$ is the same as that of $\S_V$ (see Figure \ref{fig:times}).

Though straightforward to compute, the bound in Eq.~\eqref{eq:bound} is not tight, as it is acquired by bounding the statistical error $\epsilon_{\bm v\ell}$ in the empirical estimate $\tilde T_{\bm v\ell,0}$ of $\bk{T_{\bm v\ell,0}}_\rho$ by $\bbk{\epsilon_{\bm v\ell}^2}\le\Gamma_\ell^2$.
The individual bounds on $\bbk{\epsilon_{\bm v\ell}^2}$ for each axis $\bm v$ and degree $\ell$ are tight, but these bounds cannot all be achieved simultaneously.
There is therefore still room for improvement on the bound in Eq.~\eqref{eq:bound} by maximizing $\E_V$ over the set of all physical qudit states $\rho$.
We discuss this maximization problem in Appendix \ref{sec:exact}, but leave its full solution to future work.
We also note that the reconstruction error bound in Eq.~\eqref{eq:bound} obeys the ``standard quantum limit'' of $\sim1/n$ scaling in the number of measurements.
In principle, this scaling can be improved to $\sim1/n^2$ by preparing and measuring entangled copies of many qudits \cite{giovannetti2006quantum}.

The error scales $\S_V$ and $\epsilon_V$ provide pessimistic upper bounds on statistical error, which can be calculated without prior knowledge of the true qudit state $\rho$.
The actual error in the reconstruction $\tilde\rho_V$ of a particular state $\rho$ may be considerably smaller, and may depend on $\rho$ itself.
Written out in full, the mean squared distance between $\tilde\rho_V$ and $\rho$ is (see Appendix \ref{sec:bound})
\begin{align}
  \E_V\p{\rho}^2 = \sum_{\bm v,\bm w,\ell}
  \bk{\bm v|\p{M_{V\ell}^{-1}}^\dag M_{V\ell}^{-1}|\bm w}
  \bbk{\epsilon_{\bm v\ell}\epsilon_{\bm w\ell}}.
  \label{eq:error_cls}
\end{align}
The covariances $\bbk{\epsilon_{\bm v\ell}\epsilon_{\bm w\ell}}$ are generally determined by the sources of measurement error in any given experiment, but will typically satisfy $\bbk{\epsilon_{\bm v\ell}\epsilon_{\bm w\ell}} = \delta_{\bm v\bm w}\bbk{\epsilon_{\bm v\ell}^2}$ because measurements along $\bm v$ are independent of measurements along $\bm w$.
If measurement error is limited by shot noise, then (see Appendix \ref{sec:exact})
\begin{align}
  \E_V\p{\rho}^2 \stackrel{\t{SNL}}{=} \f1n \sum_\ell
  \sp{\bk{\chi_{V\ell}|\rho_\ell}
    - \bk{\rho_\ell|\N_{V\ell}|\rho_\ell}},
  \label{eq:error_qnt}
\end{align}
where $\stackrel{\t{SNL}}{=}$ indicates equality in the ``shot-noise-limited'' regime; $\ket{\rho_\ell} \equiv \sum_m \rho_{\ell m}\ket{m}$ is a vector of the polarization operator components $\rho_{\ell m}$ of $\rho$, defined in Eq.~\eqref{eq:trans_state}; and the matrix $\N_{V\ell}$ and vector $\ket{\chi_{V\ell}}$ are defined below.
While the true shot-noise-limited error in $\tilde\rho_V$ cannot be known exactly without knowing $\rho$, this error can be estimated a posteriori by $\E_V\p{\rho}\approx\E_V\p{\tilde\rho_V}$.
After constructing an estimate $\tilde\rho_V$ of $\rho$, the complexity of computing the error $\E_V\p{\tilde\rho_V}$ from Eq.~\eqref{eq:error_qnt} is the same as that of computing $\S_V$ or $\epsilon_V$ (see Figure \ref{fig:times}).

We now define $\N_{V\ell}$ and $\ket{\chi_{V\ell}}$ for the sake of completion, but note that these definitions can be skipped without consequence for the remaining discussions in this paper.
The matrix $\N_{V\ell}$ is
\begin{align}
  \N_{V\ell} \equiv M_{V\ell}^\dag
  \diag\sp{\p{M_{V\ell}^{-1}}^\dag M_{V\ell}^{-1}} M_{V\ell},
\end{align}
where $\diag\sp{X}$ sets the off-diagonal parts of $X$ to zero.
The vector $\ket{\chi_{V\ell}}\equiv\sum_m \chi^V_{\ell m} \ket{m}$ is defined by
\begin{align}
  \chi^V_{LM} &\equiv
  \sum_\ell \obk{\N_{V\ell}|\D_M|\tilde g_{L\ell}}, \\
  \tilde g_{L\ell} &\equiv \sum_{m,m'}
  \obk{T_{L,m+m'}|T_{\ell m}^\dag T_{\ell m'}} \op{m}{m'}, \\
  \D_M &\equiv \sum_{m,m'} \delta_{M,m'-m} \op{mm'}.
\end{align}
Here $\tilde g_{L\ell}$ is essentially a matrix of structure constants for the polarization operator algebra (see Appendix \ref{sec:trans_prod}), and $\D_M$ simply picks off the $M$-th diagonal of the matrix it acts on.

\section{Tomography protocol}
\label{sec:protocol}

The ability to certify a statistical error bound on the empirical estimate $\tilde\rho_V$ of an unknown quantum state $\rho$ motivates the following protocol for spin qudit tomography:
\begin{enumerate}
\item Select a random set of measurement axes $V$ by uniformly sampling points on the sphere\footnote{To sample a point $\p{\alpha,\beta}$ from the uniform distribution on the sphere (with azimuthal angle $\alpha$ and polar angle $\beta$), you can sample a point $\p{a,b}\in[0,1]\times[0,1]$ from the uniform distribution on the unit square, and then set $\alpha=2\pi a$ and $\beta=\arccos\p{1-2b}$.}, and use any standard minimization algorithm to optimize the $2\abs{V}$ parameters in $V$ (two angles for each point $\bm v\in V$) by minimizing the quantum error scale $\epsilon_V$ in Eq.~\eqref{eq:bound}.
If $\abs{V}$ is too large for such optimization, you can simply generate many sets of random measurement axes, and then choose the set with the smallest quantum error scale $\epsilon_V$.
Note that computing the error scale $\epsilon_V$ requires, for each $\ell\in\set{0,1,\cdots,d-1}$, constructing the measurement matrix $M_{V\ell}$ in Eq.~\eqref{eq:block} and computing its singular value decomposition.
Save all measurement matrix data associated with the final measurement axes $V$ for later use.
\item For each axis $\bm v\in V$, make $n$ measurements of spin projection, and set $\tilde\Pi_{\bm v\mu}\approx\bk{\Pi_{\bm v\mu}}_\rho$ to the fraction of times in which the measurement outcome was $\mu$.
\item Use the the estimates $\tilde\Pi_{\bm v\mu}$ of $\bk{\Pi_{\bm v\mu}}_\rho$ to compute estimates of $\bk{T_{\bm v\ell,0}}_\rho$,
\begin{align}
  \tilde T_{\bm v\ell,0}
  \equiv \sum_\mu \bk{\mu|T_{\ell,0}|\mu} \tilde\Pi_{\bm v\mu},
\end{align}
where the matrix elements of $T_{\ell,0}$ are provided in Eq.~\eqref{eq:trans_op}.
\item Denoting the nonzero singular values of $M_{V\ell}$ by $M^V_{\ell k}$ and the corresponding left singular vectors by $\bm x^V_{\ell k} = \sum_{\bm v} x^V_{\ell k \bm v} \ket{\bm v}$, compute the  operators and coefficients
\begin{align}
  Q_{\ell k} &\equiv \f1{M^V_{\ell k}} \sum_{\bm v}
  \p{x^V_{\ell k\bm v}}^* T_{\bm v\ell,0}, \\
  \tilde\rho^V_{\ell k} &\equiv \f1{M^V_{\ell k}} \sum_{\bm v}
  x^V_{\ell k\bm v} \, \tilde T_{\bm v\ell,0},
\end{align}
and combine them into the estimate
\begin{align}
  \tilde\rho_V = \sum_{\ell,k} \tilde\rho^V_{\ell k} Q_{\ell k}
  \approx \rho.
\end{align}
\end{enumerate}
The expected reconstruction error in $\tilde\rho_V$, or its root-mean-square distance from $\rho$, is provided by Eq.~\eqref{eq:error_cls}.
If measurement error is shot-noise-limited, then the error in $\tilde\rho_V$ is approximately $\E_V\p{\rho}\approx\E_V\p{\tilde\rho_V}$ and can be computed from Eq.~\eqref{eq:error_qnt}.
If $\tilde\rho_V$ has negative eigenvalues, its distance from $\rho$ can be reduced with maximum-likelihood corrections \cite{smolin2012efficient}, which will additionally guarantee that $\tilde\rho_V$ satisfies all requirements for being a physical state.

The tomography protocol outlined above leaves open the question of {\it how many} measurement axes to use.
Though $2d-1$ measurement axes may be sufficient to perform full state tomography, this is not necessarily the best choice of $\abs{V}$.
Increasing the number of measurement axes generally decreases the quantum error scale $\epsilon_V$, but comes at the cost of having to estimate more observables.
At a fixed total number of measurements, increasing $\abs{V}$ reduces the number of measurements $n$ devoted to each axis $\bm v\in V$.
This trade-off begs the question: how should one choose the number of measurement axes, $\abs{V}$?

\begin{figure}
  \centering
  \includegraphics{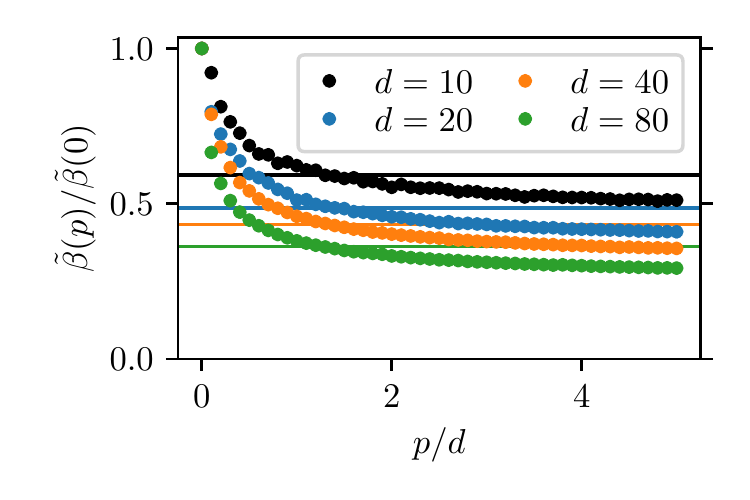}
  \caption{Empirical measurement-adjusted error scales $\tilde\beta(p)$ with $p$ excess measurement axes, determined by minimizing over $10^3$ choices of measurement axes $V$ or 5 minutes of runtime (for each $p$), whichever comes first.
    Color indicates the qudit dimension $d$.
    The rapid initial drop in $\tilde\beta(p)$ implies that using more measurement axes can substantially lower the upper bound on reconstruction error provided in Eq.~\eqref{eq:bound}, and that these benefits plateau after $p\approx d$.
    Horizontal reference lines mark the smallest measurement-adjusted error scales $\min_\theta\beta_\theta/\tilde\beta(0)$ achievable with the method in Ref.~\cite{newton1968measurability}, which is parameterized by an arbitrarily chosen angle $\theta$.}
  \label{fig:axes}
\end{figure}

The reconstruction error bound in Eq.~\eqref{eq:bound} nominally provides a straightforward answer: at a fixed total number of measurements, $N=n\abs{V}$, the number of measurement axes should be chosen to minimize the (squared) reconstruction error $\E_V\p{\rho}^2<\epsilon_V^2/n\propto\epsilon_V^2\abs{V}$.
We therefore consider the measurement-adjusted error scale $\beta(p)$ defined by
\begin{align}
  \beta(p)^2
  \equiv \min_V\Set{\epsilon_V^2\abs{V}:\abs{V}=2d-1+p},
\end{align}
where $p$ is the number of ``extra'' measurement axes exceeding $2d-1$.
Though we cannot minimize over all suitable choices of measurement axes $V$ to compute $\beta(p)$, we can compute an empirical upper bound $\tilde\beta(p)\ge\beta(p)$ by minimizing over a large number of randomly chosen $V$.
Figure \ref{fig:axes} shows the results of such empirical minimization, where we find that $\tilde\beta(p)$ drops substantially with $p$ before plateauing at $p\approx d$, after which there are only minor benefits to using more measurement axes.
In the interest of reducing experimental complexity as well as the runtime of our randomized tomography protocol, which grow linearly in $\abs{V}$, we therefore conclude that this protocol should be performed with $\abs{V}\approx3d$ measurement axes.
We provide the best measurement axes that we found for a randomized tomography protocol with $d\le30$ and $\abs{V}=3d$ in Ref.~\cite{tomo_codes}.

For reference, Figure \ref{fig:axes} also shows the smallest measurement-adjusted error scales $\beta_\theta$ achievable with the method of Ref.~\cite{newton1968measurability}, which is comparable to those achieved with our randomized protocol at $\abs{V}\approx3d$.
The method of Ref.~\cite{newton1968measurability} requires choosing an angle $\theta$, namely the polar angle of all measurement axes, and provides no prescription for making this choice.
We therefore find the optimal choice of $\theta$ by minimizing the error scale $\beta_\theta$ over all $\theta$ (see Appendix \ref{sec:angle_search}), and show $\min\beta_\theta/\tilde\beta(0)$ in Figure \ref{fig:axes}.
Empirically, we find that the optimal angle for the method of Ref.~\cite{newton1968measurability} is $\theta_{\t{opt}}\approx\frac{\pi}{2}(1-\frac1{1.34d})$ (see Appendix \ref{sec:angle_search}), which approaches $\pi/2$ as $d\to\infty$.
However, the error scale $\beta_{\pi/2}=\infty$, reflecting the fact that full state tomography is impossible with measurement axes lying in a single plane.
The method of Ref.~\cite{newton1968measurability} therefore requires extremely careful fine-tuning of measurement axis orientations for large spin dimensions.
For this reason, we expect our randomized tomography protocol be be more robust to errors in axis orientation.
We leave a detailed analysis of robustness to errors in axis orientation and the effect of these errors on state reconstruction to future work.

As a final point, we note that any information about an unknown qudit state $\rho$, obtained from prior knowledge or preliminary measurement data, can be used to construct tailored or adaptive measurement protocols \cite{huszar2012adaptive, ferrie2014selfguided, granade2016practical, pereira2018adaptive} that are more efficient in terms of the number of measurements required to estimate $\rho$ to a fixed precision.
We leave the development of tailored and adaptive measurement protocols to future work as well.

\begin{acknowledgments}
We thank Anthony M.~Polloreno and Jeremy T.~Young for helpful feedback.
This work is supported  by the DARPA DRINQs grant, the ARO single investigator award W911NF-19-1-0210, AFOSR grant FA9550-19-1-0275, NSF grant PHY-1820885, NSF grant PHY-1734006 (JILA-PFC), and by NIST.
\end{acknowledgments}

\bibliography{qudit_tomo.bib}
\onecolumngrid
\appendix

\section{Rotating polarization operators}
\label{sec:rotations}

Denoting the state of a spin-$s$ particle spin spin projection $\mu$ onto a quantization axis by $\ket{s\mu}$, we define
\begin{align}
  S_\z \equiv \sum_{\mu=-s}^s \mu \op{s\mu},
  &&
  S_\pm \equiv \sum_{\mu=-s}^s
  \sqrt{s\p{s+1}-\mu\p{\mu\pm1}} \op{s,\mu\pm1}{s\mu},
  \label{eq:spin_ops}
\end{align}
as well as
\begin{align}
  S_\x \equiv \f12\p{S_+ + S_-},
  &&
  S_\y \equiv -\f\i2\p{S_+-S_-},
  &&
  \bm S \equiv \p{S_\x,S_\y,S_\z}.
\end{align}
The spin vector $\bm S$ generates rotations of a spin-$s$ system in 3D space.
Specifically, the operator $e^{-\i\theta\bm S\cdot\uv n}$ rotates a spin-$s$ system by an angle $\theta$ about the unit vector $\uv n$.

Observing that $S_\z=T_{1,0}$ and $S_\pm\propto T_{1,\pm1}$, we can use the operator product expansion of the polarization operators (see Appendix \ref{sec:trans_prod}), the properties of Clebsch-Gordan coefficients, the properties of Wigner $6$-$j$ symbols, and a computer algebra system to simplify the commutators
\begin{align}
  \sp{S_\z,T_{\ell m}} = m\, T_{\ell m},
  &&
  \sp{S_\pm,T_{\ell m}} = \sqrt{\ell\p{\ell+1}-m\p{m\pm 1}}\, T_{\ell,m\pm1},
  \label{eq:spin_trans}
\end{align}
which implies that $T_{\ell m}$ is a spherical tensor operator, whose degree degree $\ell$ is preserved under rotations generated by $\bm S$.
Moreover, by comparing Eqs.~\eqref{eq:spin_ops} and \eqref{eq:spin_trans} we see that the polarization operators $T_{\ell m}$ transform identically to spin-$\ell$ particles under the (adjoint) action of the spin operators $S_\z$ and $S_\pm$.
For any triplet of angles $\bm\omega=\p{\alpha,\beta,\gamma}$, we can therefore define the rotation operator
\begin{align}
  R\p{\bm\omega} \equiv
  e^{-\i\alpha S_\z} e^{-\i\beta S_\y} e^{-\i\gamma S_\z},
\end{align}
and expand rotated polarization operators as
\begin{align}
  T_{\bm\omega\ell m} \equiv
  R\p{\bm\omega} T_{\ell m} R\p{\bm\omega}^\dag
  = \sum_{n=-\ell}^\ell D_{mn}^\ell\p{\bar{\bm\omega}}^* T_{\ell n},
  \label{eq:trans_rot_apndx}
\end{align}
where $\bar{\bm\omega}=\p{\gamma,\beta,\alpha}$ is the reversal of $\bm\omega$, and
\begin{align}
  D_{mn}^\ell\p{\bar{\bm\omega}}
  \equiv \bk{\ell m|R\p{\bar{\bm\omega}}|\ell n}
  = \obk{T_{\ell n} | R\p{\bm\omega}
    \otimes R\p{\bm\omega}^* |T_{\ell m}}^*
  = \obk{T_{\ell m} | R\p{-\bar{\bm\omega}}
    \otimes R\p{-\bar{\bm\omega}}^* |T_{\ell n}}
\end{align}
are matrix elements of the rotation operator $R\p{\bm\omega}$ for spin-$\ell$ particles.

For any angle doublet $\bm v=\p{\alpha,\beta}$, we define $R\p{\bm v}\equiv R\p{\alpha,\beta,0}$ and $D^\ell_{mn}\p{\bm v}=D^\ell_{mn}\p{0,\beta,\alpha}$ for shorthand.
The transformation rules in Eq.~\eqref{eq:trans_rot_apndx} imply that we can expand the phase-space representation of $T_{\ell m}$ as
\begin{align}
  T_{\ell m}^{\t{PS}}\p{\bm v}
  \equiv \bk{s_{\bm v} | T_{\ell m} | s_{\bm v}}
  = \bk{s | R\p{\bm v}^\dag T_{\ell m} R\p{\bm v} | s}
  = D^\ell_{0,m}\p{\bm v} \bk{s | T_{\ell,0} | s},
\end{align}
where
\begin{align}
  \bk{s | T_{\ell,0} | s}
  = \sqrt{\f{2\ell+1}{2s+1}} \, \bk{ss;\ell,0|ss}
  = \sqrt{\f{2\ell+1}{2s+\ell+1}
    \p{\f{\p{2s}!}{\p{2s+\ell}!}}
    \p{\f{\p{2s}!}{\p{2s-\ell}!}}},
\end{align}
and the properties of the rotation matrix elements $D^\ell_{mn}$ imply that
\begin{align}
  D^\ell_{0,m}\p{\bm v} = \sqrt{\f{4\pi}{2\ell+1}} \, Y_{\ell m}\p{\bm v},
\end{align}
so
\begin{align}
  T_{\ell m}^{\t{PS}}\p{\bm v}
  = \sqrt{\f{4\pi}{2s+\ell+1}
    \p{\f{\p{2s}!}{\p{2s+\ell}!}} \p{\f{\p{2s}!}{\p{2s-\ell}!}}} \,
  Y_{\ell m}\p{\bm v}.
\end{align}
In this way, the polarization operators are a quantum analogue of the spherical harmonics.

\section{An improved reconstruction error bound}
\label{sec:bound}

In Section \ref{sec:error} of the main text, we provided a reconstruction error bound using the assumption of Eq.~\eqref{eq:error_assumption}, namely that expectation values derived from spin projection measurements can be estimated up to uncorrelated errors with maximal variance $\epsilon^2$.
This assumption is reasonable if measurement error is dominated by experimental sources of noise, and it yields a simple derivation of the reconstruction bound in Eq.~\eqref{eq:bound_eps}.
Nonetheless, there are two problems with the assumption of Eq.~\eqref{eq:error_assumption}:
\begin{enumerate*}
\item there is no a priori guarantee for the value of $\epsilon$, which must be inferred from experimental outcomes, and
\item the assumption that all measurement errors are uncorrelated is unjustified (and generally false).
\end{enumerate*}
Here, we relax the assumption of Eq.~\eqref{eq:error_assumption} and derive an explicit error bound in terms of the qudit dimension $d$ and the number of spin projection measurements made along every measurement axis.

To this end, we fix a particular set of measurement axes $V$, and consider performing $n$ measurements of spin projection along every axis $\bm v\in V$, for a total of $N=\abs{V}\times n$ measurements.
Such a procedure is equivalent to making $N$ local measurements of the $N$-fold product state $\rho^{\otimes N}$.
For convenience, we index the tensor factors of $\rho^{\otimes N}$ by the integers $\p{i,j}$, with $i\in\set{1,2,\cdots,\abs{V}}$ specifying a measurement axis $\bm v_i\in V$, and $j\in\set{1,2,\cdots,n}$ specifying the copy of $\rho$ prepared for the $j$-th measurement spin projection along a particular axis.
We then define the projectors $\Pi_{i\mu}\equiv\op{\mu_{\bm v_i}}$ onto single-qudit states $\ket{\mu_{\bm v_i}}$ with definite spin projection $\mu$ along axis $\bm v_i\in V$, and define $\Pi_{i\mu}^j$ to be an $N$-qudit operator with $\Pi_{i\mu}$ on the $\p{i,j}$-th tensor factor and the identity elsewhere.
We denote the experimental outcome of measuring $\Pi_{i\mu}$ in the $\p{i,j}$-th copy of $\rho$ by $\tilde\Pi_{i\mu}^j\in\set{0,1}$.
In other words, $\tilde\Pi_{i\mu}^j$ is the ``single-shot estimate'' of $\Pi_{i\mu}$, with $\tilde\Pi_{i\mu}^j=1$ if outcome $\mu$ was observed on the $\p{i,j}$-th experimental trial, and $\tilde\Pi_{i\mu}^j=0$ otherwise.
An empirical estimate of the expectation value $\bk{\Pi_{i\mu}}_\rho$ is provided by the fraction of times that outcome $\mu$ was observed when measuring spin projection along axis $\bm v_i$, that is
\begin{align}
  \tilde\Pi_{i\mu} \equiv \f1n \sum_{j=1}^n \tilde\Pi_{i\mu}^j
  \approx \f1n \sum_{j=1}^n \tr\p{\rho^{\otimes N} \Pi_{i\mu}^j}
  = \tr\p{\rho \Pi_{i\mu}}.
  \label{eq:proj_estimate}
\end{align}
For reasons that will be clarified shortly, it will be useful to think of $\tilde\Pi_{i\mu}$ as an empirical estimate of $\bk{\bar\Pi_{i\mu}}_{\rho^{\otimes N}}$, where
\begin{align}
  \bar\Pi_{i\mu} \equiv \f1n \sum_{j=1}^n \Pi_{i\mu}^j
  \label{eq:mean_proj}
\end{align}
is the average of $\Pi_{i\mu}$ applied to all copies of $\rho$ for which spin projection is measured along the axis $\bm v_i$.
Eq.~\eqref{eq:proj_estimate} implies that
\begin{align}
  \tilde\Pi_{i\mu}
  \approx \bk{\bar\Pi_{i\mu}}_{\rho^{\otimes N}}
  = \bk{\Pi_{i\mu}}_\rho.
\end{align}

\subsection{Errors in the spin-projection basis}

Finite sampling error (i.e.~shot noise) generally induces statistical error $\epsilon_\O$ into the empirical estimate $\tilde\O$ of an observable $\O$:
\begin{align}
  \epsilon_\O \equiv \tilde\O - \bk{\O},
\end{align}
where the single brackets $\bk{\cdot}$ denote an expectation value with respect to the measured quantum state.
On average, this statistical error will be zero, which is to say that
\begin{align}
  \bbk{\epsilon_\O} = \bbk{\tilde\O - \bk{\O}}
  = \bk{\O - \bk{\O}}
  = 0,
\end{align}
where the double brackets $\bbk{\cdot}$ to denote statistical averaging over experimental trials that estimate $\bk{\O}$.
However, the covariance between statistical errors $\epsilon_\O$ and $\epsilon_\Q$ on the empirical estimates $\tilde\O$ and $\tilde\Q$ of observables $\O$ and $\Q$ is
\begin{align}
  \bbk{\epsilon_\O \epsilon_\Q}
  = \Bbk{\p{\tilde\O - \bk{\O}} \p{\tilde\Q - \bk{\Q}}}
  = \bk{\p{\O - \bk{\O}} \p{\Q - \bk{\Q}}}
  = \bk{\O\Q} - \bk{\O} \bk{\Q}.
\end{align}
In the context of spin qudit tomography, we can therefore define the statistical error
\begin{align}
  \epsilon_{i\mu}
  \equiv \tilde\Pi_{i\mu} - \bk{\Pi_{i\mu}}_{\rho}
  = \tilde\Pi_{i\mu} - \bk{\bar\Pi_{i\mu}}_{\rho^{\otimes N}}
\end{align}
in the empirical estimate of $\bk{\Pi_{i\mu}}_{\rho}$, and use Eq.~\eqref{eq:mean_proj} to expand
\begin{align}
  \bbk{\epsilon_{i\mu} \epsilon_{i'\mu'}}
  = \bk{\bar\Pi_{i\mu} \bar\Pi_{i'\mu'}}_{\rho^{\otimes N}}
  - \bk{\bar\Pi_{i\mu}}_{\rho^{\otimes N}}
  \bk{\bar\Pi_{i'\mu'}}_{\rho^{\otimes N}}
  = \f1{n^2} \sum_{j,j'=1}^n
  \sp{\bk{\Pi_{i\mu}^j \Pi_{i'\mu'}^{j'}}_{\rho^{\otimes N}}
    - \bk{\Pi_{i\mu}^j}_{\rho^{\otimes N}}
    \bk{\Pi_{i'\mu'}^{j'}}_{\rho^{\otimes N}}}.
  \label{eq:proj_cov_start}
\end{align}
If $\p{i,j}\ne\p{i',j'}$, then $\Pi_{i\mu}^j$ and $\Pi_{i'\mu'}^{j'}$ address different tensor factors of the product state $\rho^{\otimes N}$, so the expectation value of their product factorizes due to the fact that $\tr\sp{\p{A\otimes B}\p{A'\otimes B'}} = \tr\p{AA'}\times\tr\p{BB'}$.
This factorization can also be seen as a consequence of the fact that if $\p{i,j}\ne\p{i',j'}$, then $\Pi_{i\mu}^j$ and $\Pi_{i'\mu'}^{j'}$ are ``spatially separated'' on $\rho^{\otimes N}$, which means that their expectation values cannot have quantum correlations.
The terms in Eq.~\eqref{eq:proj_cov_start} with $\p{i,j}\ne\p{i',j'}$ therefore vanish, so
\begin{align}
  \bbk{\epsilon_{i\mu} \epsilon_{i'\mu'}}
  &= \delta_{ii'} \times \f1{n^2} \sum_{j=1}^n \sp{\bk{\Pi_{i\mu}^j \Pi_{i\mu'}^j}_{\rho^{\otimes N}}
    - \bk{\Pi_{i\mu}^j}_{\rho^{\otimes N}}
    \bk{\Pi_{i\mu'}^j}_{\rho^{\otimes N}}} \\
  &= \delta_{ii'} \times \f1n \sp{\bk{\Pi_{i\mu} \Pi_{i\mu'}}_\rho
    - \bk{\Pi_{i\mu}}_\rho \bk{\Pi_{i\mu'}}_\rho} \\
  &= \delta_{ii'} \times \f1n \cov_\rho\p{\Pi_{i\mu}, \Pi_{i\mu'}},
\end{align}
where $\cov_\rho\p{X,Y} \equiv \bk{XY}_\rho - \bk{X}_\rho \bk{Y}_\rho$.

\subsection{Errors in the polarization operator basis}

Rather than the statistical errors $\epsilon_{i\mu} \equiv \tilde\Pi_{i\mu} - \bk{\Pi_{i\mu}}_\rho$ in the estimates $\tilde\Pi_{i\mu}$ of the projectors $\Pi_{i\mu}$, we now consider the statistical errors $\epsilon_{i\ell} \equiv \tilde T_{i\ell} - \bk{T_{i\ell}}_\rho$ in the estimates $\tilde T_{i\ell}$ of the polarization operators $T_{i\ell} \equiv T_{\bm v_i\ell,0}$.
We can expand the polarization operators $T_{i\ell}$ as a sum over projectors $\Pi_{i\mu}$ as
\begin{align}
  T_{i\ell} = \sum_\mu t_{\ell\mu} \Pi_{i\mu},
  &&
  t_{\ell\mu} \equiv \bk{\mu|T_{\ell,0}|\mu}
  = \sqrt{\f{2\ell+1}{d}} \bk{s\mu;\ell,0|s\mu},
\end{align}
and likewise $\tilde T_{i\ell} \equiv \sum_\mu t_{\ell\mu} \tilde\Pi_{i\mu}$.
The covariance between errors in the polarization operator basis is then
\begin{align}
  \bbk{\epsilon_{i\ell} \epsilon_{i'\ell'}}
  = \sum_{\mu,\mu'} t_{\ell\mu} t_{\ell'\mu'}
  \bbk{\epsilon_{i\mu} \epsilon_{i'\mu'}}
  = \delta_{ii'} \times \f1n
  \sum_{\mu,\mu'} t_{\ell\mu} t_{\ell'\mu'}
  \cov_\rho\p{\Pi_{i\mu}, \Pi_{i\mu'}}
  = \delta_{ii'} \times \f1n \cov_\rho\p{T_{i\ell}, T_{i\ell'}},
\end{align}
where we used the fact that the covariance $\cov_\rho\p{X,Y}$ is linear in both $X$ and $Y$.
Due to the appearance of $\delta_{ii'}$ above and the orthogonality of polarization operators $T_{i\ell}$ and $T_{i'\ell'}$ with degrees $\ell\ne\ell'$, it turns out that only the variances $\bbk{\epsilon_{i\ell}^2}$ will ultimately contribute to reconstruction error (see Appendix \ref{sec:revisiting}).
We therefore seek to find an upper bound on $\bbk{\epsilon_{i\ell}^2}$.

To this end, we define the probability $p^i_\mu\equiv\bk{\Pi_{i\mu}}_\rho$, collect these probabilities into the classical probability distribution $p^i = \sum_\mu p^i_\mu \ket{\mu}$, and define the vector $t_\ell \equiv \sum_\mu t_{\ell\mu} \ket{\mu}$.
We then observe that
\begin{align}
  \bbk{\epsilon_{i\ell}^2} = \f1n \times \sigma_{p^i}^2\p{t_\ell},
  &&
  \sigma_p^2\p{X}
  \equiv \sum_\mu p_\mu X_\mu^2 - \p{\sum_\mu p_\mu X_\mu}^2,
\end{align}
where $\sigma_p^2\p{X}$ is the weighted variance of $X$.
This variance is maximal when $p$ has equal weight on the largest and smallest values of $X$, which implies that
\begin{align}
  \sigma_p^2\p{t_\ell} \le \Gamma_\ell^2,
  &&
  \Gamma_\ell \equiv \f{\max_\mu t_{\ell\mu} - \min_\mu t_{\ell\mu}}{2},
  &&
  \t{so}
  &&
  \bbk{\epsilon_{i\ell}^2} \le \f1n \times \Gamma_\ell^2.
\end{align}
Note that this bound on $\bbk{\epsilon_{i\ell}^2}$ is tight, as equality is achieved by the state
\begin{align}
  \rho_i^\star = \f12\p{\Pi_{i\mu_{\t{max}}} + \Pi_{i\mu_{\t{min}}}},
\end{align}
where $\mu_{\t{max}}$ ($\mu_{\t{min}}$) is the index that maximizes (minimizes) $t_{\ell\mu}$.

To find an analytical bound on $\bbk{\epsilon_{i\ell}^2}$ that is easier to interpret, we can use normalization of the polarization operators, $\obk{T_{i\ell}|T_{i\ell}}=\sum_\mu t_{\ell\mu}^2 = 1$, and the fact that all probabilities $p_\mu\le1$ to bound
\begin{align}
  \sigma_p^2\p{t_\ell} \le \sum_\mu p_\mu t_{\ell\mu}^2
  \le \sum_\mu t_{\ell\mu}^2 = 1,
  &&
  \t{so}
  &&
  \bbk{\epsilon_{i\ell}^2} < \f1n.
\end{align}
We can get a tighter bound by considering the fact that $t_{\ell\mu}^2 = t_{\ell,-\mu}^2$ due to the symmetries of the Clebsch-Gordan coefficients.
It follows that if $\mu_{\t{max}}\ne0$ then
\begin{align}
  \sigma_p^2\p{t_\ell} \le \sum_\mu p_\mu t_{\ell\mu}^2
  \le t_{\ell\mu_{\t{max}}}^2
  = \f12 \p{t_{\ell\mu_{\t{max}}}^2 + t_{\ell,-\mu_{\t{max}}}^2}
  \stackrel{\mu_{\t{max}}\ne0}{\le} \f12 \sum_\mu t_{\ell\mu}^2
  = \f12.
\end{align}
If $\mu_{\t{max}}=0$, then similarly
\begin{align}
  t_{\ell\mu_{\t{max}}}^2 + 2 t_{\ell\mu_{\t{min}}}^2
  = t_{\ell\mu_{\t{max}}}^2 + t_{\ell\mu_{\t{min}}}^2
  + t_{\ell,-\mu_{\t{min}}}^2
  \stackrel{\mu_{\t{max}}=0}{\le}
  \sum_\mu t_{\ell\mu}^2 = 1,
  &&
  \t{so}
  &&
  \abs{t_{\ell\mu_{\t{min}}}}
  \stackrel{\mu_{\t{max}}=0}{\le}
  \sqrt{\f{1 - t_{\ell\mu_{\t{max}}}^2}{2}},
\end{align}
which lets us bound
\begin{align}
  \Gamma_\ell
  = \f12 \p{t_{\ell\mu_{\t{max}}} - t_{\ell\mu_{\t{min}}}}
  \le \f12 \p{t_{\ell\mu_{\t{max}}} + \abs{t_{\ell\mu_{\t{min}}}}}
  \stackrel{\mu_{\t{max}}=0}{\le}
  \f12 t_{\ell\mu_{\t{max}}}
  + \f12\sqrt{\f{1 - t_{\ell\mu_{\t{max}}}^2}{2}}
  \equiv \lambda\p{t_{\ell\mu_{\t{max}}}}.
\end{align}
It is straightforward to show that $\lambda\p{x}$ is maximally $\lambda^\star\equiv \max_x \lambda\p{x} = \sqrt{3/8}$, so
\begin{align}
  \Gamma_\ell^2
  \stackrel{\mu_{\t{max}}=0}{\le} \p{\lambda^\star}^2
  = \f38 < \f12.
\end{align}
Altogether, we thus find that in all cases
\begin{align}
  \sigma_p^2\p{t_\ell} \le \f12,
  &&
  \t{so}
  &&
  \bbk{\epsilon_{i\ell}^2} \le \f1{2n}.
\end{align}

\subsection{Revisiting the reconstruction error bound}
\label{sec:revisiting}

We now revisit the derivation of reconstruction error in Section \ref{sec:error} to make use of the bounds on variances $\bbk{\epsilon_{i\ell}^2}$.
To recap, for a set of measurement axes $V=\set{\bm v}$ and degrees $\ell\in\set{0,1,\cdots,d-1}$ we construct the measurement matrix
\begin{align}
  M_V \equiv \sum_{\bm v,\ell} \ket{\bm v\ell} \obra{T_{\bm v\ell,0}},
\end{align}
which can be block diagonalized as
\begin{align}
  M_V U = \sum_\ell \op{\ell} \otimes M_{V\ell},
  &&
  U \equiv \sum_{\ell,m} \oket{T_{\ell m}} \bra{\ell m},
  &&
  M_{V\ell} = \sum_{m,\bm v} D^\ell_{m,0}\p{\bm v} \op{\bm v}{m},
\end{align}
where $D^\ell_{mn}\p{\bm v}\equiv\bk{\ell m|R\p{\bm v}|\ell n}$ is a (Wigner) rotation matrix element for a spin-$\ell$ particle.
The block-diagonal structure of $M_V$ allows us to index its singular values $M^V_{\ell m}$ and corresponding (normalized) left singular vectors $\bm x^V_{\ell m} = \sum_i x^V_{\ell m i} \ket{\bm v_i}$ by the indices $\p{\ell,m}$, where the integer $\abs{m}\le\ell$.
These singular vectors and values define the orthonormal operators
\begin{align}
  Q^V_{\ell m} \equiv \sum_i \p{q^V_{\ell mi}}^* T_{i\ell},
  &&
  q^V_{\ell m i} \equiv \f{x^V_{\ell m i}}{M^V_{\ell m}},
\end{align}
where $i\in\set{1,2,\cdots,\abs{V}}$ indexes an axis $\bm v_i\in V$, with $T_{i\ell} \equiv T_{\bm v_i\ell}$.
The state $\rho$ can be expanded in the basis of these operators as
\begin{align}
  \rho = \sum_{\ell,m} \bk{{Q^V_{\ell m}}^\dag}_\rho Q^V_{\ell m},
\end{align}
and the estimates $\tilde T_{i\ell}$ of $\bk{T_{i\ell}}_\rho$ can be used to construct the following estimate $\tilde\rho_V$ of $\rho$:
\begin{align}
  \tilde\rho_V \equiv \sum_{\ell,m}
  \sp{\sum_i q^V_{\ell mi} \tilde T_{i\ell}} Q^V_{\ell m}
  \approx \sum_{\ell,m}
  \sp{\sum_i q^V_{\ell mi} \bk{T_{i\ell}}_\rho} Q^V_{\ell m}
  = \sum_{\ell,m} \bk{Q^V_{\ell m}}_\rho Q^V_{\ell m}
  = \rho.
\end{align}
Recalling that $\epsilon_{i\ell} \equiv \tilde T_{i\ell} - \bk{T_{i\ell}}_\rho$, we can use orthonormality of all $Q^V_{\ell m}$ to expand the mean squared distance between $\tilde\rho_V$ and $\rho$ as
\begin{align}
  \E_V\p{\rho}^2 \equiv \Bbk{\norm{\tilde\rho_V-\rho}^2}
  = \sum_{\ell,m,i,i'} \p{q^V_{\ell mi}}^* q^V_{\ell mi'} \bbk{\epsilon_{i\ell} \epsilon_{i'\ell}}
  = \sum_{\ell,m,i} \abs{q^V_{\ell mi}}^2 \bbk{\epsilon_{i\ell}^2}
  < \f1n \sum_\ell \Gamma_\ell^2 \S_{V\ell}^2,
  \label{eq:bound_apndx}
\end{align}
where we used the fact that $\bbk{\epsilon_{i\ell}^2}\le\Gamma_\ell^2/n$, and
\begin{align}
  \sum_{m,i} \abs{q^V_{\ell mi}}^2
  = \sum_m \p{M^V_{\ell m}}^{-2}
  = \norm{M_{V\ell}^{-1}}
  = \S_{V\ell}^2.
\end{align}
Here $M_{V\ell}^{-1}$ is the left inverse of $M_{V\ell}$.
The fact that $\bbk{\epsilon_{i\ell}^2}<1/2n$ also implies that
\begin{align}
  \E_V\p{\rho}^2 < \f1{2n} \sum_\ell \S_{V\ell}^2 = \f{\S_V^2}{2n}.
\end{align}
Note that the bound in Eq.~\eqref{eq:bound_apndx} is not tight, as the individual bounds on the variances $\bbk{\epsilon_{i\ell}^2}$ cannot all be achieved simultaneously.
There is therefore still room for improvement on the bound in Eq.~\eqref{eq:bound} by maximizing $\E_V$ over the set of physically achievable qudit states $\rho$.

\section{Exact reconstruction error}
\label{sec:exact}

Here we find exact expressions for reconstruction error, which can be used to estimate the error in a given reconstruction $\tilde\rho_V$ of an unknown state $\rho$ after performing tomography.
To this end, we start with Eq.~\eqref{eq:bound_apndx} from Appendix \ref{sec:revisiting} to write
\begin{align}
  \E_V\p{\rho}^2
  = \sum_{\ell,m,i,i'} \p{q^V_{\ell mi}}^* q^V_{\ell mi'} \bbk{\epsilon_{i\ell}\epsilon_{i'\ell}}
  = \f1n \sum_{\ell,i} \abs{\tilde{\bm q}_{\ell i}}^2 \cov_\rho\p{T_{i\ell},T_{i\ell}},
  &&
  \abs{\tilde{\bm q}_{\ell i}}^2 = \sum_m \abs{q_{\ell mi}}^2,
  \label{eq:error_SM_start}
\end{align}
where $\tilde{\bm q}_{\ell i} = \sum_m \p{q_{\ell mi}}^* \ket{m}$, and we used the fact that $\bbk{\epsilon_{i\ell}\epsilon_{i'\ell}} = \delta_{ii'}\times\cov_\rho\p{T_{i\ell},T_{i\ell}}/n$.
Identifying the singular value decomposition $M_{V\ell} = U_{V\ell} \Sigma_{V\ell} W_{V\ell}^\dag$, we then we observe that $\tilde{\bm q}_{\ell i} = \Sigma_{V\ell}^{-1} U_{V\ell}^\dag \ket{\bm v_i}$, which allows us to simplify
\begin{align}
  \abs{\tilde{\bm q}_{\ell i}}^2
  = \bk{\bm v_i | U_{V\ell} \Sigma_{V\ell}^{-2} U_{V\ell}^\dag | \bm v_i}
  = \bk{\bm v_i| \p{M_{V\ell}^{-1}}^\dag M_{V\ell}^{-1} | \bm v_i}.
\end{align}
Using the fact that all $T_{i\ell}=T_{i\ell}^\dag$, we can also expand
\begin{align}
  \cov_\rho\p{T_{i\ell},T_{i\ell}}
  = \cov_\rho\p{T_{i\ell}^\dag,T_{i\ell}}
  = \sum_{m,m'} D^\ell_{0,m}\p{\bm v_i} D^\ell_{0,m'}\p{\bm v_i}^*
  \cov_\rho\p{T_{\ell m}^\dag, T_{\ell m'}},
\end{align}
which implies that
\begin{align}
  \E_V\p{\rho}^2
  = \f1n \sum_{\ell,i,m,m'}
  D^\ell_{0,m'}\p{\bm v_i}^* \abs{\tilde{\bm q}_{\ell i}}^2
  D^\ell_{0,m}\p{\bm v_i}
  \cov_\rho\p{T_{\ell m}^\dag, T_{\ell m'}}.
\end{align}
Altogether, this reconstruction error can be expressed more compactly by defining the {\it covariance matrix}
\begin{align}
  \C_\ell\sp{\rho} \equiv
  \sum_{m,m'} \cov_\rho\p{T_{\ell m}^\dag, T_{\ell m'}} \op{m}{m'},
\end{align}
and the {\it noise matrix}
\begin{align}
  \N_{V\ell} \equiv M_{V\ell}^\dag \diag\sp{\p{M_{V\ell}^{-1}}^\dag
    M_{V\ell}^{-1}} M_{V\ell},
\end{align}
where $\diag\sp{X}$ sets all off-diagonal entries of $X$ to zero, in terms of which
\begin{align}
  \E_V\p{\rho}^2 = \f1n \sum_\ell \obk{\N_{V\ell}|\C_\ell\sp{\rho}},
  \label{eq:error_SM_mid}
\end{align}
where $\obk{X|Y}=\tr\p{X^\dag Y}$ is a trace inner product.

The result in Eq.~\eqref{eq:error_SM_mid} essentially expresses reconstruction error as a weighted sum of the covariances $\cov_\rho\p{T_{\ell m},T_{\ell m'}}$, where the weights are given by the corresponding matrix elements of the noise matrix $\N_{V\ell}$.
This expression is perhaps the most physically meaningful form of the reconstruction error $\E_V\p{\rho}$ that we will consider in this work, but in practice it turns out that Eq.~\eqref{eq:error_SM_mid} is inconvenient and inefficient to evaluate for any given state $\rho$.
To find a more practical expression of reconstruction error, we use the fact that
\begin{align}
  \bk{T_{\ell m}^\dag}_\rho
  = \obk{\rho|T_{\ell m}^\dag}
  = \tr\p{\rho T_{\ell m}^\dag}
  = \tr\p{T_{\ell m}^\dag \rho}
  = \obk{T_{\ell m}|\rho},
\end{align}
to expand the covariance matrix as
\begin{align}
  \C_\ell\sp{\rho}
  &= \sum_{m,m'} \op{m}{m'} \sp{\obk{\rho|T_{\ell m}^\dag T_{\ell m'}}
    - \obk{\rho|T_{\ell m}^\dag} \obk{\rho|T_{\ell m'}}} \\
  &= \sum_{m,m'} \op{m}{m'} \sp{\obk{T_{\ell m'}^\dag T_{\ell m}|\rho}
    - \obk{T_{\ell m}|\rho} \obk{T_{\ell m'}^\dag|\rho}} \\
  &= \sum_{m,m'} \op{m}{m'}\I \sp{\obk{T_{\ell m'} T_{\ell m}|\rho}
    - \obk{T_{\ell m}|\rho} \obk{T_{\ell m'}|\rho}}
\end{align}
where we define the inversion operator $\I\equiv\sum_m \p{-1}^m\op{-m}{m}$.
We then expand the product $T_{\ell m'} T_{\ell m}$ as
\begin{align}
  \obk{T_{\ell m'} T_{\ell m}|\rho}
  = \sum_L g_{\ell m'm}^L \obk{T_{L,m'+m}|\rho},
  &&
  g_{\ell m'm}^L \equiv \obk{T_{L,m'+m} | T_{\ell m'} T_{\ell m}}
  = f_{\ell m';\ell m}^{L,m'+m},
\end{align}
where the (real) factors $f_{\ell m';\ell m}^{L,m'+m}$ are provided in Appendix \ref{sec:trans_prod}.
Substituting the covariance matrix back into Eq.~\eqref{eq:error_SM_mid} and replacing $\obk{T_{\ell m}|\rho}\to\rho_{\ell m}$, we get
\begin{align}
  \E_V\p{\rho}^2
  = \f1n \sp{\sum_{\ell,m} \p{\chi^V_{\ell m}}^* \rho_{\ell m}
  - \sum_{\ell,m,m'} \bk{m'|\I\N_{V\ell}|m}
  \rho_{\ell m} \rho_{\ell m'}},
  \label{eq:error_comps}
\end{align}
where
\begin{align}
  \chi^V_{LM}
  &\equiv \sum_{\ell,m,m'} \delta_{M,m'+m}
  \bk{m'|\I\N_{V\ell}|m}^* g^L_{\ell m'm} \\
  &= \sum_{\ell,m,m'} \delta_{M,-m'+m}
  \bk{m|\N_{V\ell}|m'} \p{-1}^{m'} g^L_{\ell,-m',m} \\
  &= \sum_\ell \obk{\N_{V\ell}|\D_M|\I g_{L\ell}}
\end{align}
can be written in terms of the matrices
\begin{align}
  g_{L\ell} \equiv \sum_{m,m'} g^L_{\ell m'm} \op{m'}{m},
  &&
  \D_M \equiv \sum_{m,m'} \delta_{M,-m'+m} \op{m'm}.
\end{align}
Here $\D_M$ simply picks off the $M$-th diagonal of the matrix it acts on, such that $\obk{\N_{V\ell}|\D_M|\I g_{L\ell}}$ is an inner product of the $M$-th diagonal of $\I g_{L\ell}$ with the $\p{-M}$-th diagonal of $\N_{V\ell}$.
Defining the $\p{2\ell+1}$-component vectors
\begin{align}
  \ket{\rho_\ell} \equiv \sum_m \rho_{\ell m} \ket{m},
  &&
  \ket{\chi_{V\ell}} \equiv \sum_{\ell,m} \chi^V_{\ell m} \ket{m},
\end{align}
we can write the expansion in Eq.~\eqref{eq:error_comps} in the vectorized form
\begin{align}
  \E_V\p{\rho}^2
  = \f1n \sum_\ell \sp{\bk{\chi_{V\ell}|\rho_\ell}
    - \bk{\rho_\ell|\N_{V\ell}|\rho_\ell}}.
  \label{eq:error_SM}
\end{align}

\subsection*{Comments on a tight reconstruction error bound}

In principle, maximizing the reconstruction error in Eq.~\eqref{eq:error_SM} over all qudit states $\rho$ would provide a tight upper bound on reconstruction error for any set of axes $V$.
To simplify this task somewhat, we first maximize Eq.~\eqref{eq:error_SM} over all $\rho$ with $\tr\p\rho=1$: this maximum occurs at a ``state'' $\sigma_V^\star$ whose components are given by
\begin{align}
  \ket{\sigma_{V\ell}^\star} \stackrel{\ell\ne0}{\equiv}
  \f12 \N_{V\ell}^{-1} \ket{\chi_{V\ell}},
  &&
  \ket{\sigma_{V,0}^\star} \equiv \f1{\sqrt{d}} \ket{0}.
\end{align}
The corresponding maximum of $\E_V$ is given by
\begin{align}
  \E_V\p{\sigma_V^\star}^2 = \f1n \sum_{\ell>0}
  \sp{\f14 \bk{\chi_{V\ell}|\N_{V\ell}^{-1}|\chi_{V\ell}}
  - \f1d \tr\p{\N_{V\ell}}},
\end{align}
where the $\tr\p{\N_{V\ell}}$ terms above come from simplifying the $\ell=0$ terms of Eq.~\eqref{eq:error_SM} with $\rho\to\sigma_V^\star$.
While $\E_V\p{\sigma_V^\star}$ is a strict upper bound on $\E_V\p{\rho}$ over all $\rho$ with $\tr\p{\rho}=1$, this bound turns out to be useless in practice, because $\sigma_V^\star$ will generally be a non-physical ``state'' with negative eigenvalues.
To find tight bound on $\E_V\p{\rho}$ over the space of physical qudit states $\rho$, we also need to constrain $\rho$ to have no negative eigenvalues.
Equipped with $\sigma_V^\star$ and $\E_V\p{\sigma_V^\star}$, we can expand
\begin{align}
  \E_V\p{\rho}^2 = \E_V\p{\sigma_V^\star}^2
  - \f1n \norm{\rho-\sigma_V^\star}_V^2,
  &&
  \norm{X}_V^2 \equiv \sum_\ell \bk{X_\ell|\N_{V\ell}|X_\ell},
\end{align}
where $X_\ell\equiv\sum_m\obk{T_{\ell m}|X}\ket{m}$ is a vector of the degree-$\ell$ components of $X$ in the polarization operator basis, and $\norm{X}_V$ is a noise-weighted norm of $X$.
Maximizing $\E_V$ over all qudit states $\rho$ thus amounts to finding the closest physical qudit state $\rho$ to $\sigma_V^\star$, with distance measured by the metric $D_V\p{X,Y}\equiv\norm{X-Y}_V$.
We leave this minimization problem to future work, and note that solving it will likely require making use of the positivity conditions derived in Ref.~\cite{kryszewski2006positivity}.
A loose lower bound on $\norm{\rho-\sigma_V^\star}_V$ can be found by minimization under the constraint $\norm{\rho}\le1$, which may provide a tighter upper bound on $\E_V\p{\rho}$ than that in Eq.~\eqref{eq:bound} of the main text.

\section{Polarization operator product expansion}
\label{sec:trans_prod}

The polarization operators on the $d$-dimensional Hilbert space of a spin-$s$ system (with $s\equiv\frac{d-1}{2}$) are defined by
\begin{align}
  T_{\ell m} \equiv \sqrt{\f{2\ell+1}{2s+1}} \sum_{\mu,\nu=-s}^s
  \bk{s\mu;\ell m|s\nu} \op{\nu}{\mu},
\end{align}
where $\bk{s\mu;\ell m|s\nu}$ is a Clebsh-Gordan coefficient that enforces $\ell\in\set{0,1,\cdots,2s}$ and $m\in\set{-\ell,-\ell+1,\cdots,\ell}$.
We wish to compute the coefficients of the operator product expansion
\begin{align}
  T_{\ell_1 m_1} T_{\ell_2 m_2}
  = \sum_{L,M} f_{\ell_1 m_1;\ell_2 m_2}^{LM} T_{LM},
  &&
  f_{\ell_1 m_1;\ell_2 m_2}^{LM}
  \equiv \obk{T_{LM} | T_{\ell_1 m_1} T_{\ell_2 m_2}},
\end{align}
which allow us to simplify the commutators in Eq.~\eqref{eq:spin_trans} of Appendix \ref{sec:rotations}.
Using the symmetry properties of Clebsch-Gordan coefficients, namely
\begin{align}
  \bk{\ell_1 m_1; \ell_2 m_2| L M}
  &= \p{-1}^{\ell_2+m_2} \sqrt{\f{2L+1}{2\ell_1+1}}
  \bk{L,-M; \ell_2 m_2| \ell_1,-m_1} \\
  \bk{\ell_1 m_1; \ell_2 m_2| L M}
  &= \p{-1}^{\ell_1+\ell_2-L}
  \bk{\ell_1,-m_1; \ell_2,-m_2| L,-M},
\end{align}
we can find that the polarization operators transform under conjugation as
\begin{align}
  T_{\ell m}^\dag
  = \sqrt{\f{2\ell+1}{2s+1}}
  \sum_{\mu,\nu} \p{-1}^m \bk{s\nu;\ell,-m|s\mu} \op{\mu}{\nu}
  = \p{-1}^m T_{\ell,-m},
\end{align}
which implies that
\begin{align}
  f_{\ell_1 m_1;\ell_2 m_2}^{LM}
  = \p{-1}^M \sqrt{\f{\p{2L+1}\p{2\ell_1+1}\p{2\ell_2+1}}
    {\p{2s+1}\p{2s+1}\p{2s+1}}}
  \sum_{\mu,\nu,\rho} \bk{s\nu;L,-M|s\mu}
  \bk{s\rho;\ell_1m_1|s\nu} \bk{s\mu;\ell_2m_2|s\rho}.
\end{align}
Replacing Clebsch-Gordan coefficients by Wigner 3-$j$ symbols with the identity
\begin{align}
  \bk{\ell_1 m_1; \ell_2 m_2| L M}
  = \p{-1}^{2\ell_2} \p{-1}^{L-M} \sqrt{2L+1}
  \begin{pmatrix}
    L & \ell_2 & \ell_1 \\
    -M & m_2 & m_1
  \end{pmatrix},
\end{align}
we can use the fact that $2\ell_2$ is always even (because $\ell_2$ is always an integer) to expand
\begin{multline}
  f_{\ell_1 m_1;\ell_2 m_2}^{LM}
  = \p{-1}^M \sqrt{\p{2L+1}\p{2\ell_1+1}\p{2\ell_2+1}} \\
  \times \sum_{\mu,\nu,\rho} \p{-1}^{3s-\mu-\nu-\rho}
  \begin{pmatrix}
    s & L & s \\
    -\mu & -M & \nu
  \end{pmatrix}
  \begin{pmatrix}
    s & \ell_1 & s \\
    -\nu & m_1 & \rho
  \end{pmatrix}
  \begin{pmatrix}
    s & \ell_2 & s \\
    -\rho & m_2 & \mu
  \end{pmatrix}.
\end{multline}
This sum can be simplified by the introduction of Wigner 6-$j$ symbols, giving us
\begin{align}
  f_{\ell_1 m_1;\ell_2 m_2}^{LM}
  &= \p{-1}^{2s+M} \sqrt{\p{2L+1}\p{2\ell_1+1}\p{2\ell_2+1}}
  \begin{pmatrix}
    L & \ell_1 & \ell_2 \\
    M & -m_1 & -m_2
  \end{pmatrix}
  \begin{Bmatrix}
    L & \ell_1 & \ell_2 \\
    s & s & s
  \end{Bmatrix} \\
  &= \p{-1}^{2s+L} \sqrt{\p{2\ell_1+1}\p{2\ell_2+1}}
  \bk{\ell_1 m_1; \ell_2 m_2| LM}
  \begin{Bmatrix}
    \ell_1 & \ell_2 & L \\
    s & s & s
  \end{Bmatrix}.
\end{align}

\section{Optimizing the method of Newton and Young}
\label{sec:angle_search}

\begin{figure}
  \centering
  \includegraphics{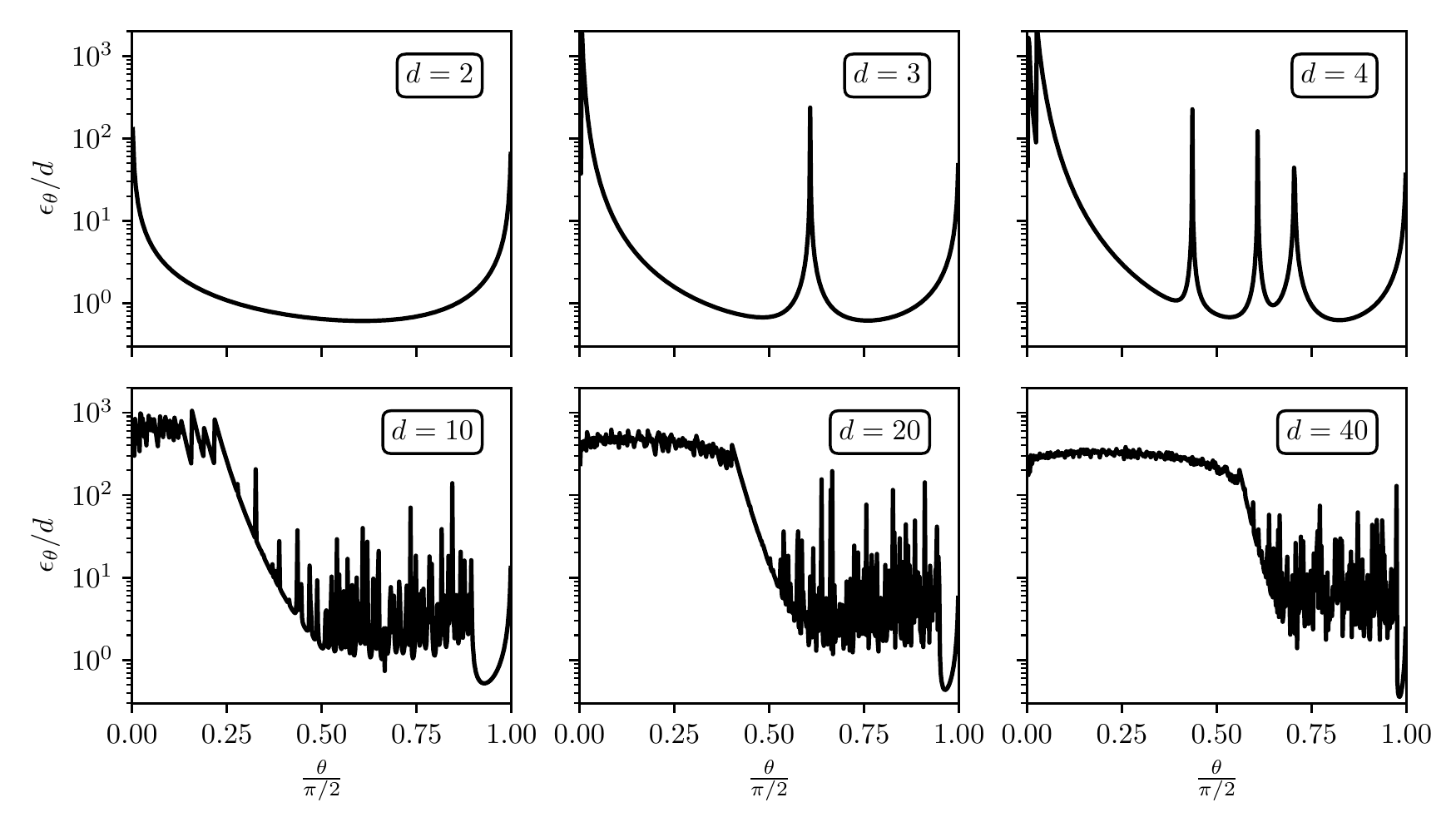}
  \caption{Quantum error scale $\epsilon_\theta$ as a function of the polar angle $\theta$ in the tomography method of Ref.~\cite{newton1968measurability} for a few qudit dimensions $d$.}
  \label{fig:angle_sweep}
\end{figure}

\begin{figure}
  \centering
  \includegraphics{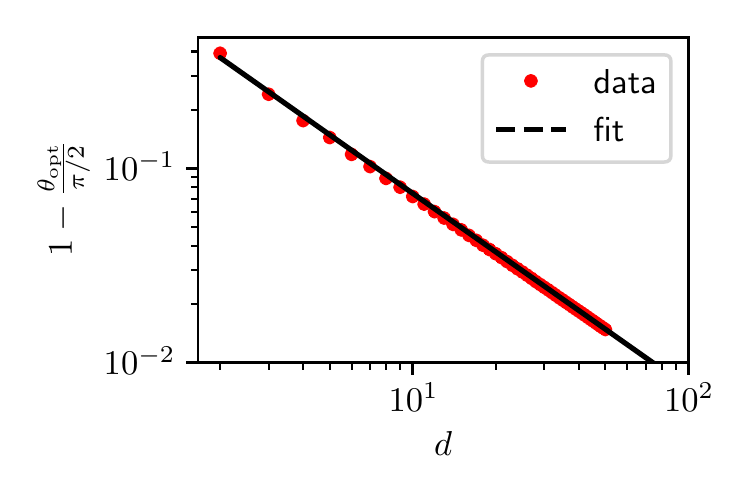}
  \caption{Optimum angle $\theta_{\t{opt}}$ as a function of qudit dimension $d$ for the tomography method of Ref.~\cite{newton1968measurability}, and a fit to $\theta_{\t{opt}}=\frac{\pi}{2}(1-\frac1{xd})$ finding $x\approx1.34$.}
  \label{fig:opt_angles}
\end{figure}

Ref.~\cite{newton1968measurability} constructs an explicit protocol for spin qudit tomography, which involves measuring spin projection along $2d-1$ axes equally spaced at a polar angle $\theta$.
However, this method does not provide any prescription for choosing $\theta$.
Here, we show the importance of making a good choice of $\theta$, and empirically find the optimal value of $\theta_{\t{opt}}$ that minimizes the corresponding quantum error scale $\epsilon_\theta$, which controls state reconstruction error.
To this end, Figure \ref{fig:angle_sweep} shows the quantum error scale $\epsilon_\theta$ as a function of the polar angle $\theta$ in the tomography method of Ref.~\cite{newton1968measurability} for a few qudit dimensions $d$.
While a good choice of $\theta$ yields an error scale $\epsilon_\theta\approx d$ (for the dimensions shown), this error scale can increase by orders of magnitude for poor choices of $\theta$.
In turn, Figure \ref{fig:opt_angles} shows the optimal angle $\theta_{\t{opt}}$ as a function of the qudit dimension $d$, together with a fit to $\theta_{\t{opt}}=\frac{\pi}{2}(1-\frac1{xd})$ finding $x\approx1.34$.

\end{document}